\documentclass[journal]{IEEEtran}
\IEEEoverridecommandlockouts
\usepackage{graphicx}
\usepackage{amsmath,amssymb, amsfonts, amsthm}
\usepackage{multirow}
\usepackage[ruled,lined,commentsnumbered]{algorithm2e}
\usepackage{caption}
\usepackage{color}
\usepackage{shuffle}
\usepackage{makecell}

\newtheorem{theorem}{Theorem}
\newtheorem{definition}[theorem]{Definition}
\newtheorem{corollary}[theorem]{Corollary}
\newtheorem{example}[theorem]{Example}
\newtheorem{remark}[theorem]{Remark} 
\newtheorem{proposition}[theorem]{Proposition}
\newtheorem{lemma}[theorem]{Lemma}

\title{\LARGE \bf
Automatic Generation of Optimal Reductions of Distributions
} 
\author{Liyong Lin, Tom\'a\v s Masopust, W. Murray Wonham, Rong Su%
  \thanks{L.~Lin and R.~Su are affiliated with School of Electrical and Electronic Engineering, Nanyang Technological University. T.~Masopust is affiliated with Institute of Mathematics, Czech Academy of Sciences. W.~M.~Wonham is affiliated with Department of Electrical and Computer Engineering, University of Toronto. This work is supported by Natural Sciences and Engineering Research Council (NSERC) Grant Number \#DG\_480599, RVO 67985840, the ATMRI-CAAS project on Air Traffic Flow Management with the project reference number ATMRI: 2014-R7-SU and the National Research Foundation of Singapore Delta-NTU Corporate Lab Program with the project reference of DELTA-NTU CORP-SMA-RP2. Email addresses: liyong.lin@utoronto.ca, masopust@math.cas.cz, wonham@control.utoronto.ca, rsu@ntu.edu.sg.}}

\begin{document}
\maketitle
\thispagestyle{empty}
\pagestyle{empty}

\begin{abstract}
A reduction of a source distribution is a collection of smaller sized distributions that are collectively equivalent to the source distribution with respect to the property of decomposability. That is, an arbitrary language is decomposable with respect to the source distribution if and only if it is decomposable with respect to each smaller sized distribution (in the reduction). The notion of reduction of distributions has previously been proposed to improve the complexity of decomposability verification. In this work, we address the problem of generating (optimal) reductions of distributions automatically. A (partial) solution to this problem is provided, which consists of 1) an incremental algorithm for the production of candidate reductions and 2) a reduction validation procedure. In the incremental production stage, backtracking is applied whenever a candidate reduction that cannot be validated is produced. A strengthened substitution-based proof technique is used for reduction validation, while a fixed template of candidate counter examples is used for reduction refutation; put together, they constitute our (partial) solution to the reduction verification problem. In addition, we show that a recursive approach for the generation of (small) reductions is easily supported.

\end{abstract}
\vspace{-2pt}
\section{Introduction}
A language $L \subseteq \Sigma^*$ is said to be {\em decomposable} with respect
to a distribution, i.e., a tuple of non-empty sub-alphabets, of $\Sigma$ if $L$ is equal to the synchronous product of its projections onto
the respective sub-alphabets. The notion of decomposability is
well studied and applied in the literature on supervisory control
theory, see for example~\cite{WH91, LW88, KMS12, CW10, RW92, KM13}, and synthesis of
synchronous products of transition systems~\cite{A06}. As a special case of co-observability~\cite{KM13}, the property of decomposability is already quite expensive to verify in general~\cite{WH91, KMS12}. Indeed, in general, it is quite unlikely that a polynomial time algorithm would exist since the problem of decomposability verification is PSPACE-complete, even when $L$ is prefix-closed (Theorem 4.24 in~\cite{A06}).
 A basic problem of interest is to identify tractable fragments of the decomposability verification problem, based on structures of the distributions. The first such result is given in~\cite{KMS12} for the verification of conditional decomposability, i.e., decomposability with a rather restricted structure imposed on the distributions. There are also some techniques that exploit both structures of the distributions and prefix-closeness of the test languages. The earliest such result that we know of appears in~\cite{WH91}, which has been generalized in the Appendix of~\cite{LWSW17b}. We shall not require prefix-closeness of the test languages in this work, to exploit solely the structures of distributions.

Recently, a notion of reduction of distributions~\cite{LWSW17b},~\cite{LWSW17a} has been proposed to improve the complexity of decomposability verification and support the parallel verification of decomposability, which is heavily inspired by (and generalizes) the result on polynomial time verification of conditional decomposability in~\cite{KMS12}. Intuitively, a reduction of a given (source) distribution is a set of smaller sized distributions satisfying the property that an arbitrary language $L$ is decomposable with respect to the source distribution if and only if it is decomposable with respect to each smaller sized distribution in the reduction. If such a reduction exists, then we can verify the decomposability of $L$ with respect to every smaller sized distribution (in the reduction) to determine the decomposability of $L$ with respect to the source distribution. This divide-and-conquer approach allows us to reduce the verification complexity for a large class of distributions~\cite{KMS12},~\cite{LWSW17b}. Several basic challenges encountered in this approach include the problem of efficient production of candidate reductions and the problem of reduction verification. It is of considerable interest for us to find algorithmic solutions to these two basic problems, which can be used for automatic generation of (optimal) reductions. Optimal reductions of distributions then correspond to optimal reduction of verification complexity in our approach\footnote{This statement assumes that reduction computation is not the bottleneck, which is often the case. In practice, we can run direct verification (w.r.t. $\Delta$) and the reduction-based verification in parallel and output the first answer that becomes available.}.

The main motivation 
for our study on the decomposability verification problem is its
close relationship with the problem of existence of decentralized supervisors\footnote{The decomposability verification problem is a special case of the problem of existence of decentralized supervisors~\cite{LWSW17b}: an instance of the first problem with prefix-closed test language $L \subseteq \Sigma^*$ and distribution $\Delta=(\Sigma_i)_{i=1}^n$ is an instance of the second problem with specification $L$, decentralized control architecture $(A_i)_{i=1}^n$, where $A_i=(\Sigma_i, \Sigma_i)$, and plant $G$, where $L_m(G)=L(G)=\Sigma^*$ (see Proposition 43 in~\cite{LWSW17b} for more details).} (see for example~\cite{RW92, KM13, LWSW17b}). In particular, a structural approach for improving
the complexity of decomposability verification can lead to
a better understanding of the complexity for deciding the existence of decentralized supervisors, in relation to the
structures of the decentralized control architectures~\cite{LWSW17b},~\cite{LWSW17a}. A similar line of research for understanding the decidability of the decentralized supervisor synthesis problem appears in~\cite{LSSWS14} and~\cite{LSS16}, which is based on some central structural results in trace theory~\cite{S92}. The problem of reduction verification is related to the problem of finding decision-theoretic reduction functions (in descriptive complexity)~\cite{CIM10},~\cite{LWSW17b}. Indeed, verifying a candidate reduction of a distribution is equivalent to verifying the identity function to be a reduction (function) between two decision problems (of fixed format) about decomposability~\cite{LWSW17b}. In general, however, even the problem of verifying whether the identity function is a reduction between two given decision problems is undecidable~\cite{JK13}. It is of interest to know whether the same conclusion also holds on our special problem setup\footnote{We conjecture that the problem of reduction verification is decidable.}. 

Some basic results have been established in~\cite{LWSW17b}. In this work, we continue to develop some results to address the following technical problems listed in~\cite{LWSW17b}, with a particular emphasis on Technical Problem 3:
\begin{enumerate}
\item Given any distribution $\Delta$ and any set $P$ of distributions, determine whether $P$ is a reduction of $\Delta$ 
\item Given any distribution $\Delta$ of $\Sigma$, determine whether it has a reduction
\item Given any distribution $\Delta$, compute an optimal reduction, if there exists a reduction
\item Given any set $P$ of distributions, determine the existence of a distribution $\Delta$ such that $P$ is a reduction of $\Delta$
\item Given any distribution $\Delta$ of $\Sigma$ and any reduction $P$ of $\Delta$, determine whether $P$ is compact, that is, whether $P$ involves no redundancy
\end{enumerate}

The contributions of this study, compared with~\cite{LWSW17b}, are the following.
\begin{itemize}
\item We show that the two partially ordered sets studied in~\cite{LWSW17b}, i.e., the set of all distributions of an alphabet $\Sigma$ and the set of candidate reductions of a distribution $\Delta$, are indeed finite lattices, thus complete. Then, we can talk about the least upper bounds and the greatest lower bounds for sets of distributions and sets of candidate reductions.
\item We then show that a set $P$ of distributions is a reduction of $\Delta$ only if $\Delta$ is the greatest lower bound of $P$. It then follows that Technical Problem 4 is reducible to Technical Problem 1; thus, we only need to call the reduction verification procedure for once to solve Technical Problem 4. We also show that the above order-theoretic necessary condition is equivalent to the necessary condition stated in Lemma 25 of~\cite{LWSW17b} (reproduced as Lemma~\ref{lemma: c} in this work), which is based on the notion of generalized dependence relation.
\item We implement a reduction verification procedure, which combines a strengthened version of the  substitution-based (automated) proof technique~\cite{LWSW17b} for reduction validation and the candidate counter examples based techniques for reduction refutation. Thus, given any candidate reduction $P$ of $\Delta$, if $\Delta$ can be obtained by using the strengthened substitution-based automated proof, then $P$ is a reduction of $\Delta$; and a fixed template of candidate counter examples is used to refute $P$, which turns out to be quite effective in refuting candidate reductions that are not valid. 
\item Structural results are also developed for verifying the decomposability of the fixed template of candidate counter examples, which can be used to show that some classes of distributions (e.g., distributions of ring-like structures) have no reductions. 
\item We develop an incremental algorithm for the production of candidate reductions, which is guided by Lemma 24 and Lemma 25 of~\cite{LWSW17b} (Lemmas~\ref{prop: nece} and~\ref{lemma: c} in this work). It is tailored for the efficient production of a subclass of candidate reductions that are meet-consistent\footnote{The property of meet-consistency shall be discussed later. It is a necessary condition for a candidate reduction to be a valid reduction of a distribution.}. Backtracking will be enforced whenever a (meet-consistent) candidate reduction that cannot be validated is produced. The idea is to first explore small candidate reductions, i.e., candidate reductions of small width. Thus, it works quite effectively for those distributions which have small reductions. Since small candidate reductions often correspond to those high up elements\footnote{We have the Hasse diagram, where the top element lies in the top and the bottom element lies in the bottom. The high up elements lie close to the top element and the low down elements lie close to the bottom element.} in the lattice of the set of candidate reductions, this incremental production algorithm generates the candidate reductions from the higher up elements to those lower down ones in the lattice. 
\item For those distributions that do not have small reductions, the same reduction validation procedure may potentially be called for many times before a reduction is computed, using the above incremental production algorithm (with backtracking). To address this problem, we also develop a ``conservative" recursive algorithm for computing reductions (guided by Theorem 22 of~\cite{LWSW17b}, which is reproduced as Theorem~\ref{thm: exiten} in this work), by generating lower down candidate reductions in the lattice first. Once a candidate reduction is validated, it can be recursively improved by computing reductions of the constituent distributions, if the reductions exist.
\end{itemize}
While the decidability statuses of the main technical problems still remain unknown, we believe that this work constitutes a major step, compared with~\cite{LWSW17b}, towards practically solving the problems of reduction verification, determining the existence of reductions and automatic generation of optimal reductions. We have carried out a number of experiments on determining the existence of reductions. And the results obtained are quite encouraging (see Appendix B for a small list of examples), and we have not found examples where the reduction verification procedure fails to conclude\footnote{The reduction verification problem is a central problem in the investigation of the decidability. We would like to know whether the reduction verification procedure is complete. We experiment on the easier problem of determining the existence of reductions, which can be reduced to the reduction verification problem~\cite{LWSW17b}.}.

The rest of this paper will be organized as follows. Section~\ref{sec: notation} is devoted to
preliminaries. In Section~\ref{sec: br}, we provide several basic results. In Section~\ref{section:RVd}, we then present the reduction verification procedure, which consists of both reduction validation and reduction refutation. Section~\ref{sec:agr} then addresses the problem of automatic generation of small or  (near-)optimal reductions, where the central aim is to achieve efficient production of candidate reductions. We then draw the conclusions
in Section~\ref{sec: condu}. Some additional material and experimental results are included in the Appendices.

\vspace{-4pt}
\section{Preliminaries}
\label{sec: notation}
We assume that the reader is familiar with basic theories
of formal languages and finite automata~\cite{HU79}. Some knowledge about order
theory~\cite{DP02, WMW17} is also assumed. In the following, additional notation and
terminology are introduced, including some definitions and results from~\cite{LWSW17b}.

Let $[1,n]$ denote the set $\{1, 2, \ldots, n\}$. For any two sets $A$ and $B$, we write $A-B$ to denote the set-theoretic difference of $A$ and $B$. The cardinality of any set $A$ is denoted by $|A|$. For any given alphabet $\Sigma$, a  \emph{distribution} of $\Sigma$ of size $n$ is an $n$-tuple $\Delta=(\Sigma_1,\Sigma_2,\ldots,\Sigma_n)$ of non-empty sub-alphabets of $\Sigma$ such that $\Sigma=\bigcup_{i=1}^{n}\Sigma_i$ and the sub-alphabets are pairwise incomparable with respect to set inclusion. We sometimes also view the distribution
$\Delta$ as the set $\{\Sigma_1, \Sigma_2, \ldots, \Sigma_n\}$ for convenience and use $|\Delta|$ to denote the size of $\Delta$. Given a distribution $\Delta=(\Sigma_1,\Sigma_2,\ldots,\Sigma_n)$ of $\Sigma$, we have $n$ projections $P_i$ from $\Sigma^{*}$ to $\Sigma_{i}^{*}$ and $n$ inverse projections $P_i^{-1}$ from $\Sigma_i^{*}$ to $2^{\Sigma^*}$. Also, both projections and inverse projections are naturally extended to the mappings between languages. The \emph{synchronous product} $\lVert_{i=1}^{n}L_i$ of languages $ L_i $ over $\Sigma_i$ is defined as $\bigcap_{i=1}^{n}P_{i}^{-1}(L_i)$. A language $L \subseteq \Sigma^*$ is said to be {\em decomposable} with respect to distribution $\Delta=(\Sigma_1, \Sigma_2, \ldots, \Sigma_n)$ of $\Sigma$ if $L=\lVert_{i=1}^nP_i(L)$. The
straightforward approach for checking the decomposability of $L$ with respect to $\Delta$ is of worst case time complexity\footnote{Whenever we talk about verification complexity, we implicitly assume that $L$ is regular. The general development of this work need not require $L$ to be regular.} $\mathcal{O}(((n+1)|\Sigma|-\sum_{i=1}^n |\Sigma_i|)m^{n+1})$, where $m$ is the state size of the given deterministic finite automaton that recognizes $L$. The {\em decomposition closure} $\lVert_{i=1}^nP_i(L)$ of $L$ with respect to $\Delta$ is denoted by $L^{\Delta}$. Let 
\vspace{-4pt}
\begin{center}
$s_1 \shuffle s_2=\{v_1u_1v_2u_2\ldots v_ku_k \mid s_1=v_1v_2\ldots v_k, s_2=u_1u_2\ldots u_k, \textit{where } v_i, u_i \in \Sigma^* \textit{ for each } i\in [1, k]\}$
\end{center}
\vspace{-4pt}
 denote the \emph{shuffle} of any two strings $s_1, s_2 \in \Sigma^*$. For example, the shuffle of strings $s_1=abb$ and $s_2=a$ is the language $s_1 \shuffle s_2=\{aabb, abab, abba\}$.

\begin{figure}[t]
\centering
\hspace*{-1mm}
\includegraphics[width=1.4in, height=0.6in]{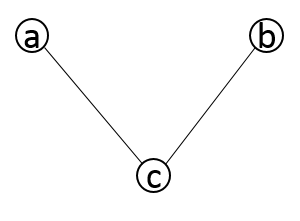} 
\captionsetup{justification=centering}
\caption{visualization of $I_{\Delta}$ induced by $\Delta=(\{a, b\}, \{c\})$}
\label{fig:inde}
\end{figure}

Each distribution $\Delta=(\Sigma_1, \Sigma_2, \ldots, \Sigma_n)$ induces a {\em dependence relation} $D_{\Delta} \subseteq \Sigma \times \Sigma$, which is defined below.
\vspace{-4pt}
\begin{center}
$D_{\Delta}:=\{(a,b) \in \Sigma \times \Sigma \mid \exists i \in [1,n], a,b \in \Sigma_i\}$
\end{center}
\vspace{-4pt}
Clearly, $D_{\Delta}$ is reflexive and symmetric. Then, the complement $I_{\Delta}:=\Sigma \times \Sigma -D_{\Delta}$ of $D_{\Delta}$ is said to be the {\em independence relation} induced by $\Delta$, which is irreflexive and symmetric. For any $(a, b) \in I_{\Delta}$, $a$ and $b$ are said to be {\em independent symbols}. We can use an undirected graph $(\Sigma, I_{\Delta})$ for the visualization of an independence relation. For example, the independence relation $I_{\Delta}=\{(a, c), (c, a), (b, c), (c, b)\}$ induced by the distribution $\Delta=(\{a, b\}, \{c\})$ of $\Sigma=\{a, b, c\}$ can be visualized in Fig.~\ref{fig:inde}. We may also use the notation $I_{\Delta}=\{\{a, c\}, \{b, c\}\}$, where, for example, $\{a, c\}$ is used to represent both $(a, c)$ and $(c, a)$. Let $\Sigma'$ be any non-empty sub-alphabet of $\Sigma$. We define $\Sigma' \sqsubseteq \Delta$ if $\exists i \in [1, n], \Sigma' \subseteq \Sigma_i$. Clearly, for any sub-alphabet $\Sigma' \subseteq \Sigma$ with cardinality 1, $\Sigma' \sqsubseteq \Delta$ holds for any distribution $\Delta$ of $\Sigma$. Apparently, for any two symbols $a, b \in \Sigma$, $\{a, b\} \sqsubseteq \Delta$ if and only if $(a, b) \in D_{\Delta}$. Thus, $\sqsubseteq$ naturally generalizes the notion of a dependence relation.

Two strings $s, s'$ over $\Sigma$ are said to be {\em trace equivalent} with respect to the independence relation $I_{\Delta}$, denoted by $s \sim_{I_{\Delta}} s'$, if there exist strings $v_0, \ldots, v_k$ such that $s=v_0, s'=v_k$ and for each $i \in [1, k]$, there exist some $u_i, u'_i \in \Sigma^*$ and $a_i, b_i \in \Sigma$ such that $(a_i, b_i) \in I_{\Delta}$, $v_{i-1}=u_ia_ib_iu'_i$ and $v_i=u_ib_ia_iu'_i$. Intuitively, two strings are trace equivalent if each one of them
can be obtained from the other by a sequence of permutations
of adjacent symbols that are independent. And the set of trace equivalent strings of $s$ for $I_{\Delta}$ is said to be the {\em trace closure} of $s$ with respect to $I_{\Delta}$, denoted by $[s]_{I_{\Delta}}$. The {\em trace closure} $[L]_{I_{\Delta}}$ of a
language $L \subseteq \Sigma^*$ is defined to be the set $\bigcup_{s \in L}[s]_{I_{\Delta}}$. A language $L$ is said to be {\em trace-closed} with respect to $I_{\Delta}$ if $L=[L]_{I_{\Delta}}$. For example, consider the independence relation $I_{\Delta}=\{(a, c), (c, a), (b, c), (c, b)\}$ induced by $\Delta=(\{a, b\}, \{c\})$. The strings $bac, cba$ are trace equivalent with respect to $I_{\Delta}$ since $bac \sim_{I_{\Delta}} bca \sim_{I_{\Delta}} cba$.  $L=\{bac, cba, bca\}$ is trace-closed with respect to $I_{\Delta}$ since we have $[L]_{I_{\Delta}}=L=\{bac, cba, bca\}$. We shall often identify each singleton with the unique element it contains.

We recall the following well-established result (see Proposition 4.3 in~\cite{F07}, where a detailed proof is given).
\begin{lemma}
\label{lemm: separation}
Let $L_i \subseteq \Sigma_i^*$ for $i \in [1, n]$ and $\Sigma=\bigcup_{i=1}^n\Sigma_i$. If $\bigcup_{i \neq j}(\Sigma_i \cap \Sigma_j) \subseteq \Sigma_0 \subseteq \Sigma$, then $P_0(\lVert_{i=1}^n L_i)=\lVert_{i=1}^n P_0(L_i)$. Here, $P_0: \Sigma^* \rightarrow \Sigma_0^*$ is the natural projection and $P_0(L_i) \subseteq (\Sigma_0 \cap \Sigma_i)^*$ is viewed as a language over $\Sigma_0 \cap \Sigma_i$.
\end{lemma}

In the rest of this section, we shall introduce some useful definitions and results from~\cite{LWSW17b} to make this paper mostly self-contained. The most important definition is the definition of a reduction of a distribution, which is recalled in the following.

\begin{definition}[\cite{LWSW17b}]
\label{def:reduction}
A set $\{\Delta_1, \Delta_2, \ldots, \Delta_l\}$ of distributions of $\Sigma$, where $l \geq 2$, is said to be a reduction of distribution $\Delta$ if 1) an arbitrary language $L \subseteq \Sigma^*$ is decomposable with respect to $\Delta_i$ for each $i \in [1, l]$ if and only if it is decomposable with respect to $\Delta$; 2) for each $i \in [1, l], |\Delta_i|<|\Delta|$.
\end{definition}
A distribution $\Delta$ of $\Sigma$ of size $n \geq 3$ is said to be {\em reducible} if it has a reduction. It is said to be
$(l, k)$-{\em reducible} if it has a reduction $P=\{\Delta_1, \Delta_2, \ldots, \Delta_l\}$, where
$\max_{i \in [1, l]}|\Delta_i|=k$; and the reduction $P$ will be referred to as
an $(l, k)$-{\em reduction} of $\Delta$, in which $l$ is called the {\em height}, $k$ the
{\em width}, and $(l, k)$ the {\em dimension}. The width is the main determining factor for the worst case time complexity, and the height of a reduction corresponds to the number of parallel processors needed in the parallel verification. A
reduction of $\Delta$ is said to be {\em compact} if none of its proper subsets
is also a reduction of $\Delta$. Formally, we say that a reduction $P$ of distribution $\Delta$ is
{\em optimal}\footnote{The definition for the optimality of reductions used in this work is slightly different from that used in~\cite{LWSW17b}. In this work, we attempt to minimize both the width and the height of a reduction, while minimizing the width is the priority.} if for any other reduction $P'$ of $\Delta$, it holds that either 1) the width of $P'$ is greater than the width of $P$, or 2) if the width of $P'$ is equal to the width of $P$, then the height of $P'$ is greater than or equal to the height of $P$. Clearly, an optimal reduction is necessarily compact. The next example~\cite{LWSW17b} shows that it is possible to have two or more compact reductions of the same width, but of different heights. Thus, a compact reduction is not necessarily optimal.
\begin{example}
\label{examp: red}
Consider the distribution $\Delta=(\Sigma_1, \Sigma_2, \Sigma_3, \Sigma_4)$ of $\{a, b, c, d, e, f\}$, where $\Sigma_1=\{a, b\}$, $\Sigma_2=\{b, c\}$, $\Sigma_3=\{d, e\}$ and $\Sigma_4=\{e, f\}$. It is known that $\{\Delta_1', \Delta_2', \Delta_3'\}$ is a compact $(3, 2)$-reduction of $\Delta$~\cite{LWSW17b}, where
\[   \left\{
\begin{array}{ll}
      \Delta_1'=(\{a, b\}, \{b, c, d, e, f\}) \\
      \Delta_2'=(\{a, b, c\}, \{d, e, f\}) \\
      \Delta_3'=(\{d, e\}, \{a, b, c, e, f\}) \\
\end{array} 
\right. \]
\vspace{-4pt}
and $\{\Delta_1'', \Delta_2''\}$ is a compact $(2, 2)$-reduction of $\Delta$, where
\[   \left\{
\begin{array}{ll}
      \Delta_1''=(\{a, b, e, f\}, \{b, c, d, e\}) \\
      \Delta_2''=(\{a, b, c\}, \{d, e, f\}) \\
\end{array} 
\right. \]   
$\{\Delta_1'', \Delta_2''\}$ is an optimal reduction of $\Delta$ of dimension $(2, 2)$. Then, we can conclude that $\{\Delta_1', \Delta_2', \Delta_3'\}$ is not an optimal reduction of $\Delta$. 
\end{example}

A partial ordering $\leq_{\Sigma}$ on the set $\Delta(\Sigma)$ of all distributions of $\Sigma$ (lifting the partial ordering $\subseteq$) is defined as follows. 
\begin{definition}[$\leq_{\Sigma}$~\cite{LWSW17b}]
\label{def: order}
Given any two distributions $\Delta$ and $\Delta'$ of $\Sigma$, we define $\Delta \leq_{\Sigma} \Delta'$ if $\forall \Sigma_i \in \Delta, \exists \Sigma_j' \in \Delta', \Sigma_i \subseteq \Sigma_j'$.
\end{definition}

We have the following lemma. 
\begin{lemma}[\cite{LWSW17b}]
\label{lemma:good}
Let $\Delta$ and $\Delta'$ be any two distributions of $\Sigma$. Then, $\Delta \leq_{\Sigma} \Delta'$ if and only if for an arbitrary language $L \subseteq \Sigma^*$, the decomposability of $L$ with respect to $\Delta$ implies the decomposability of $L$  with respect to $\Delta'$.  
\end{lemma}

The next corollary follows from Lemma~\ref{lemma:good}.
\begin{corollary}[\cite{LWSW17b}]
\label{corollary: ne}
Let $\{\Delta_1, \Delta_2,\ldots, \Delta_l\}$ be any reduction of $\Delta$. Then, for each $i \in [1, l]$, we have $\Delta \leq_{\Sigma} \Delta_i$.
\end{corollary}

A distribution $\Delta'=(\Sigma_1', \Sigma_2', \ldots, \Sigma_m')$ is said to be {\em merged} from $\Delta$ if $\Delta'$ can be obtained from $\Delta=(\Sigma_1, \Sigma_2, \ldots, \Sigma_n)$ by 1) merging the sub-alphabets according to a proper partition of $[1, n]$, and followed by 2) keeping only those maximal sub-alphabets in the resulting tuple~\cite{LWSW17b}. This is illustrated by the next example~\cite{LWSW17b}.
\begin{example}
Consider the alphabet $\Sigma=\{a, b, c, d, e\}$ and the distribution $\Delta=(\Sigma_1, \Sigma_2, \Sigma_3, \Sigma_4)$ of $\Sigma$, where $\Sigma_1=\{a, b\}$, $\Sigma_2=\{b, c\}$, $\Sigma_3=\{c, d\}$ and $\Sigma_4=\{d, e\}$. If we merge $\Sigma_1$ and $\Sigma_3$, which corresponds to the partition $\{\{1, 3\}, \{2\}, \{4\}\}$ of $[1, 4]$. then we obtain the tuple $(\Sigma_1 \cup \Sigma_3, \Sigma_2, \Sigma_4)$. By keeping only those maximal sub-alphabets, we obtain the distribution $\Delta'=(\Sigma_1 \cup \Sigma_3, \Sigma_4)$ of $\Sigma$. Then, $\Delta'$ is merged from $\Delta$, by the above definition. For convenience, we say $\Delta'$ is obtained from $\Delta$ by merging $\Sigma_1$ and $\Sigma_3$. And the partition is denoted by $P_{\Delta}(\Delta')$, that is, $P_{\Delta}(\Delta')=\{\{1, 3\}, \{2\}, \{4\}\}$. 
\end{example}
 Let $\mathcal{M}(\Delta)$ denote the set of distributions that are merged from $\Delta$. We only need to consider those candidate reductions in which the distributions all belong to $\mathcal{M}(\Delta)$ and are pairwise $\leq_{\Sigma}$-incomparable~\cite{LWSW17b}. Here, two distributions $\Delta$, $\Delta'$ are $\leq_{\Sigma}$-incomparable if $\Delta \not\leq_{\Sigma} \Delta'$ and $\Delta' \not\leq_{\Sigma} \Delta$. Then, let the search space for candidate reductions of $\Delta$ be 
\vspace{-4pt}
\begin{center}
$S_{\Delta}:=\{P \in 2^{\mathcal{M}(\Delta)} \mid \forall \Delta', \Delta'' \in P, (\Delta' \neq \Delta'' \Longrightarrow \Delta' \not\leq_{\Sigma} \Delta'')\}$
\end{center}
\vspace{-4pt}
Compared with~\cite{LWSW17b}, $\varnothing$ is included in $S_{\Delta}$ here for the purpose of technical convenience (see Theorem~\ref{propo: sfin}). 

The following partial ordering $\leq_{\Delta}$ on $S_{\Delta}$ (lifting the partial ordering $\leq_{\Sigma}$) is defined.
\begin{definition}[$\leq_{\Delta}$~\cite{LWSW17b}]
Given any two elements $P, P'$ in $S_{\Delta}$, we define $P \leq_{\Delta} P'$ if $\forall \Delta_j' \in  P', \exists \Delta_i \in P, \Delta_i \leq_{\Sigma} \Delta_j'$.
\end{definition}

We assume $\mathcal{M}(\Delta)$ is non-empty, since otherwise $\Delta$ cannot have a reduction. In the partial order $(S_{\Delta}, \leq_{\Delta})$, the bottom\footnote{This claim can be verified. It is also an easy consequence of Theorem~\ref{propo: sfin}, which will be shown later. } element exists and is indeed equal to $[{\mathcal{M}(\Delta)}] \in S_{\Delta}$~\cite{LWSW17b}. Here, we have used the notation that, for any set $P$ of distributions, $[P]$ stands for the set of distributions that is obtained from $P$ by keeping only those minimal distributions (in terms of the partial ordering $\leq_{\Sigma}$). A distribution $\Delta'$ is said to be {\em minimally merged} from $\Delta$ if it can be obtained from $\Delta$ by merging two sub-alphabets $\Sigma_i$ and $\Sigma_j$ of $\Delta$, where $i \neq j$. And let $\bot(\Delta) \subseteq \mathcal{M}(\Delta)$ denote the set of minimally merged distributions from $\Delta$. $[{\bot(\Delta)}]$ is also the bottom element of $S_{\Delta}$, since $[{\mathcal{M}(\Delta)}]=[\bot(\Delta)]$~\cite{LWSW17b}. 

The next two lemmas provide necessary conditions for a set (in $S_{\Delta}$) of distributions to be a reduction of $\Delta$. Lemma~\ref{lemma: c} is a generalization of Lemma~\ref{prop: nece}, since $\sqsubseteq$ naturally generalizes the notion of a dependence relation.

\begin{lemma}[\cite{LWSW17b}]
\label{prop: nece}
If $\{\Delta_1, \Delta_2, \ldots, \Delta_l\}$ is a reduction of $\Delta$, then $I_{\Delta}=\bigcup_{i \in [1, l]}I_{\Delta_i}$ or, equivalently, $D_{\Delta}=\bigcap_{i \in [1, l]}D_{\Delta_i}$.
\end{lemma}

\begin{lemma}[\cite{LWSW17b}]
\label{lemma: c}
Let $\Delta$ be any distribution of $\Sigma$ and let $\Delta_i$ be any distribution of $\Sigma$ for each $i\in [1, l]$. If $\{\Delta_1, \Delta_2, \ldots, \Delta_l\}$ is a reduction of $\Delta$, then for each subset $\Sigma'$ of $\Sigma$ of cardinality at least 2, $(\forall i \in [1, l], \Sigma' \sqsubseteq \Delta_i) \Longleftrightarrow \Sigma' \sqsubseteq \Delta$.
\end{lemma}

The following lemma states that the set of reductions of any given distribution $\Delta$ within $S_{\Delta}$ is indeed a downward closed set with respect to the partial ordering $\leq_{\Delta}$.
\begin{lemma}[\cite{LWSW17b}]
\label{prop:down}
Let $P$ and $Q$ be any two elements of $S_{\Delta}$. If $P \leq_{\Delta} Q$ and $Q$ is a reduction of $\Delta$, then $P$ is also a reduction of $\Delta$.
\end{lemma}

Since $[{\bot(\Delta)}]$ is the bottom element of $S_{\Delta}$, we then have the next result, by Lemma~\ref{prop:down}.
\begin{theorem}[\cite{LWSW17b}]
\label{thm: exiten}
 Let $\Delta$ be any distribution of $\Sigma$. Then, $\Delta$ has a reduction if and only if $[{\bot(\Delta)}]$ is a reduction of $\Delta$. 
\end{theorem}
Thus, to determine the existence of a reduction for any given distribution $\Delta$, we only need to check whether the set $[{\bot(\Delta)}]$ is a reduction (of $\Delta$). That is, Technical Problem 2 is reduced to Technical Problem 1, i.e., we only need to call the reduction verification procedure for once.

\begin{example}[\cite{LWSW17b}]
\label{example: sf}
For the distribution $\Delta$ in Example~\ref{examp: red}, we have $\bot(\Delta)=\{\Delta_{34}, \Delta_{12}, \Delta_{13}, \Delta_{14}, \Delta_{23}, \Delta_{24}\}$, where
\[   \left\{
\begin{array}{ll}
      \Delta_{34}:=(\{a, b\}, \{b, c\}, \{d, e, f\}) \\
      \Delta_{12}:=(\{d, e\}, \{e, f\}, \{a, b, c\}) \\
      \Delta_{13}:=(\{a, b, d, e\}, \{b, c\}, \{e, f\}) \\
      \Delta_{14}:=(\{a, b, e, f\}, \{b, c\}, \{d, e\}) \\
      \Delta_{23}:=(\{b, c, d, e\}, \{a, b\}, \{e, f\}) \\
      \Delta_{24}:=(\{b, c, e, f\}, \{a, b\}, \{d, e\}) \\
\end{array} 
\right. \]
Here $\Delta_{ij}$ denotes the distribution that is obtained from $\Delta$ by merging $\Sigma_i$ and $\Sigma_j$. In this example, distributions in $\bot(\Delta)$ are $\leq_{\Sigma}$-incomparable. Theorem~\ref{thm: exiten} implies that the distribution $\Delta$ in Example~\ref{examp: red} has a reduction if and only if $[\bot(\Delta)]=\bot(\Delta)$ shown above is a reduction.
\end{example}

\vspace{-10pt} 
\section{Some Basic Results}
\label{sec: br}
In this section, we shall show that $(\Delta(\Sigma), \leq_{\Sigma})$ and $(S_{\Delta}, \leq_{\Delta})$ are indeed finite lattices, thus complete. We shall also provide an order-theoretic necessary condition for a set of distributions to be a reduction (of a source distribution), which will make it obvious that Technical Problem 4 can be reduced to Technical Problem\footnote{We shall recall the following:
\begin{itemize}
\item Technical Problem 1: Given any distribution $\Delta$ and any set $P$ of distributions, determine whether $P$ is a reduction of $\Delta$; 
\item Technical Problem 4: Given any set $P$ of distributions, determine the existence of a distribution $\Delta$ such that $P$ is a reduction of $\Delta$
\end{itemize}} 1, i.e., we only need to call the reduction verification procedure for once.
\vspace{-4pt}
\subsection{Ordering on Distributions}
It is not difficult to check that the relation $\leq_{\Sigma}$ is a partial ordering on $\Delta(\Sigma)$.  Indeed, we shall now show that $(\Delta(\Sigma), \leq_{\Sigma})$ is a finite lattice.
\begin{theorem}
\label{prop: finitelattice}
$(\Delta(\Sigma), \leq_{\Sigma})$ is a finite lattice.
\end{theorem}
{\em Proof}: 
It is not difficult to see that $(\Delta(\Sigma), \leq_{\Sigma})$ is a partial order. In the partial order $(\Delta(\Sigma), \leq_{\Sigma})$, 
\begin{enumerate}
\item the meet $\Delta \wedge \Delta'$ of any two distributions $\Delta, \Delta'$ always exists, and it is defined to be the distribution $\Delta''$ that is obtained from the set $\{\Sigma_i \cap \Sigma_j' \mid \Sigma_i \in \Delta, \Sigma_j' \in \Delta'\}$ by keeping only those maximal sub-alphabets. We now show the following.
\begin{enumerate}
\item $\Delta'' \leq_{\Sigma} \Delta$ and $\Delta'' \leq_{\Sigma} \Delta'$: for any $\Sigma_i \cap \Sigma_j' \in \Delta''$, since $\Sigma_i \cap \Sigma_j' \subseteq \Sigma_i \in \Delta$, we have $\Delta'' \leq_{\Sigma} \Delta$; and similarly, $\Delta'' \leq_{\Sigma} \Delta'$.
\item for any distribution $\Delta'''$, if $\Delta''' \leq_{\Sigma} \Delta$ and $\Delta''' \leq_{\Sigma} \Delta'$, then $\Delta''' \leq_{\Sigma} \Delta''$: Suppose that $\Delta''' \leq_{\Sigma} \Delta$ and $\Delta''' \leq_{\Sigma} \Delta'$, then for any $\Sigma_k''' \in \Delta'''$, there exist some $\Sigma_i \in \Delta$ and $\Sigma_j' \in \Delta'$ such that $\Sigma_k''' \subseteq \Sigma_i$ and $\Sigma_k''' \subseteq \Sigma_j'$. Thus, $\Sigma_k''' \subseteq \Sigma_i \cap \Sigma_j'$. If $\Sigma_i \cap \Sigma_j \in \Delta''$, then we have $\Sigma_k''' \subseteq \Sigma_i \cap \Sigma_j' \in \Delta''$; otherwise, by the definition of $\Delta''$, there exist $\Sigma_{i_1} \in \Delta, \Sigma_{j_1} \in \Delta'$ such that $\Sigma_i \cap \Sigma_j' \subset \Sigma_{i_1} \cap \Sigma_{j_1}' \in \Delta''$ ($\Sigma_i \cap \Sigma_j' \notin \Delta''$, thus it must have been removed due to the presence of some proper superset in $\Delta''$), and thus $\Sigma_k''' \subset \Sigma_{i_1} \cap \Sigma_{j_1}' \in \Delta''$. Thus, we have $\Delta''' \leq_{\Sigma} \Delta''$.
\end{enumerate}
\item the join $\Delta \vee \Delta'$ of any two distributions $\Delta, \Delta'$ always exists, and it is defined to be the distribution $\Delta''$ whose components are exactly the maximal elements (in terms of set inclusion) of $\Delta \cup \Delta'$. We now show the following.
\begin{enumerate}
\item $\Delta \leq_{\Sigma} \Delta''$ and $\Delta' \leq_{\Sigma} \Delta''$: this is straightforward from the definition of $\Delta''$ and the definition of $\leq_{\Sigma}$. Indeed, for any $\Sigma_i \in \Delta$,  if $\Sigma_i \in \Delta''$, then we have $\Sigma_i \subseteq \Sigma_i \in \Delta''$; otherwise, by the definition of $\Delta''$, there exists some $\Sigma_j'' \in \Delta''$ such that $\Sigma_i \subset \Sigma_j'' \in \Delta''$ ($\Sigma_i \notin \Delta''$, it must have been removed due to the presence of some proper superset in $\Delta''$). Thus, we have $\Delta \leq_{\Sigma} \Delta''$. Similarly, $\Delta' \leq_{\Sigma} \Delta''$.
\item for any distribution $\Delta'''$, if $\Delta \leq_{\Sigma} \Delta'''$ and $\Delta' \leq_{\Sigma} \Delta'''$, then $\Delta'' \leq_{\Sigma} \Delta'''$: this is straightforward from the definition of $\Delta''$ and the definition of $\leq_{\Sigma}$. For any $\Sigma_i'' \in \Delta''$, by the definition of $\Delta''$, $\Sigma_i'' \in \Delta$ or $\Sigma_i'' \in \Delta'$. If $\Sigma_i'' \in \Delta$, then there exists $\Sigma_m''' \in \Delta'''$ such that $\Sigma_i'' \subseteq \Sigma_m'' \in \Delta'''$, since $\Delta \leq_{\Sigma} \Delta'''$ holds. Similarly, if $\Sigma_i'' \in \Delta'$, then there exists $\Sigma_k''' \in \Delta'''$ such that $\Sigma_i'' \subseteq \Sigma_k'' \in \Delta'''$, since $\Delta' \leq_{\Sigma} \Delta'''$.  Thus, we have $\Delta'' \leq_{\Sigma} \Delta'''$. \hfill $\blacksquare$\\
\end{enumerate}
\end{enumerate}

We have the next result, which strengthens Corollary~\ref{corollary: ne}: $\Delta$ not only is a lower bound of its reduction $\{\Delta_1, \Delta_2, \ldots, \Delta_l\}$, but it is also the greatest lower bound.
\begin{proposition}
\label{theorem: char}
Let $\{\Delta_1, \Delta_2,\ldots, \Delta_l\}$ be any reduction of $\Delta$. We have $\Delta=\bigwedge_{i=1}^l \Delta_i$.
\end{proposition}
{\em Proof}: By Theorem~\ref{prop: finitelattice}, $\bigwedge_{i=1}^l \Delta_i$ exists. By Corollary~\ref{corollary: ne}, we have $\Delta \leq \bigwedge_{i=1}^l \Delta_i$, since $\{\Delta_1, \Delta_2,\ldots, \Delta_l\}$ is a reduction of $\Delta$. For any language $L \subseteq \Sigma^*$ that is decomposable with respect to distribution $\bigwedge_{i=1}^l \Delta_i$, we conclude that $L$ is also decomposable with respect to each $\Delta_j$, by Lemma~\ref{lemma:good}, since $\bigwedge_{i=1}^l \Delta_i \leq_{\Sigma} \Delta_j$ for each $j \in [1, l]$. Since $\{\Delta_1, \Delta_2,\ldots, \Delta_l\}$ is a reduction of $\Delta$, we then conclude that $L$ is decomposable with respect to $\Delta$. That is, for any language $L \subseteq \Sigma^*$ that is decomposable with respect to $\bigwedge_{i=1}^l \Delta_i$,  $L$ is decomposable with respect to $\Delta$. By Lemma~\ref{lemma:good}, we then have $\bigwedge_{i=1}^l \Delta_i \leq_{\Sigma} \Delta$. Thus, $\Delta=\bigwedge_{i=1}^l \Delta_i$.
\hfill $\blacksquare$\\

Proposition~\ref{theorem: char} provides another necessary condition for a set of distributions to be a reduction of a source distribution $\Delta$. It is of interest to compare which one is stronger, Lemma~\ref{lemma: c} or Proposition~\ref{theorem: char}. Indeed, they are equivalent.
\begin{theorem}
\label{theore: equi}
Let $\Delta$ be any distribution of $\Sigma$ and let $\Delta_i$ be any distribution of $\Sigma$, for each $i\in [1, l]$. $\Delta=\bigwedge_{i=1}^l \Delta_i$ if and only if for each subset $\Sigma'$ of $\Sigma$ of cardinality at least 2, $(\forall i \in [1, l], \Sigma' \sqsubseteq \Delta_i) \Longleftrightarrow \Sigma' \sqsubseteq \Delta$.
\end{theorem}

{\em Proof}: Suppose $\Delta=\bigwedge_{i=1}^l \Delta_i$. Let $\Sigma'$ be any subset of $\Sigma$ with $|\Sigma'| \geq 2$. Suppose for each $i \in [1, l], \Sigma' \sqsubseteq \Delta_i$. Then we have 
$\Sigma' \sqsubseteq \bigwedge_{i=1}^l \Delta_i=\Delta$, by the definitions of $\sqsubseteq$ and the meet operation $\wedge$. The direction $\Sigma' \sqsubseteq \Delta \Longrightarrow (\forall i \in [1, l], \Sigma' \sqsubseteq \Delta_i)$ is automatic from $\Delta=\bigwedge_{i=1}^l \Delta_i \leq_{\Sigma} \Delta_j$, for each $j \in [1, l]$. On the other hand, suppose that for each subset $\Sigma'$ of $\Sigma$ of cardinality at least 2, $(\forall i \in [1, l], \Sigma' \sqsubseteq \Delta_i) \Longleftrightarrow \Sigma' \sqsubseteq \Delta$. We shall first show that $\Delta \leq_{\Sigma} \Delta_i$ for each $i \in [1, l]$. This is not difficult. Indeed, for any $\Sigma' \in \Delta$, we have $\Sigma' \sqsubseteq \Delta$ by definition. We conclude\footnote{If $|\Sigma'| \geq 2$, then $\Sigma' \sqsubseteq \Delta_i$ follows from $\Sigma' \sqsubseteq \Delta$ and the supposition $(\forall i \in [1, l], \Sigma' \sqsubseteq \Delta_i) \Longleftrightarrow \Sigma' \sqsubseteq \Delta$; if $|\Sigma'|=1$, $\Sigma' \sqsubseteq \Delta_i$ holds trivially.} that $\Sigma' \sqsubseteq \Delta_i$ for each $i \in [1, l]$, i.e., for each $\Sigma' \in \Delta$, there exists $\Sigma'' \in \Delta_i$ such that $\Sigma' \subseteq \Sigma''$. Thus, we have that $\Delta \leq_{\Sigma} \bigwedge_{i=1}^l \Delta_i$. For any $\Sigma' \in \bigwedge_{i=1}^l \Delta_i$, by the definitions of $\wedge$ and $\sqsubseteq$, we have $\Sigma' \sqsubseteq \Delta_i$ for each $i \in [1, l]$. Thus, we conclude\footnote{If $|\Sigma'| \geq 2$, then $\Sigma' \sqsubseteq \Delta$ follows from $\Sigma' \sqsubseteq \Delta_i$, for each $i \in [1, l]$, and the supposition $(\forall i \in [1, l], \Sigma' \sqsubseteq \Delta_i) \Longleftrightarrow \Sigma' \sqsubseteq \Delta$; if $|\Sigma'|=1$, $\Sigma' \sqsubseteq \Delta$ holds trivially.} that $\Sigma' \sqsubseteq \Delta$. That is, for any $\Sigma' \in \bigwedge_{i=1}^l \Delta_i$, there exists $\Sigma'' \in \Delta$ such that $\Sigma' \subseteq \Sigma''$. We have $\bigwedge_{i=1}^l \Delta_i \leq_{\Sigma} \Delta$. Thus, $\Delta=\bigwedge_{i=1}^l \Delta_i$.
 \hfill $\blacksquare$. \\

Another immediate consequence of Proposition~\ref{theorem: char} is that a set of distributions can only be the reduction of their greatest lower bound in the lattice $(\Delta(\Sigma), \leq_{\Sigma})$. Thus, Technical Problem 4 can be reduced to Technical Problem 1 in the following way: $\{\Delta_1, \Delta_2,\ldots, \Delta_l\}$ is a reduction of some distribution if and only if $\{\Delta_1, \Delta_2,\ldots, \Delta_l\}$ is a reduction of $\bigwedge_{i=1}^l \Delta_i$. 

The necessary condition given in Lemma~\ref{lemma: c} is essentially order-theoretic, as has been shown in Theorem~\ref{theore: equi}. It turns out that it is also possible to reformulate the necessary condition provided in Lemma~\ref{prop: nece}. The key is the following.  
\begin{lemma}
\label{lemma: indered}
For any set $\{\Delta_1, \Delta_2, \ldots, \Delta_l\}$ of distributions of $\Sigma$, we have $\bigcup_{i \in [1, l]}I_{\Delta_i}=I_{\bigwedge_{i=1}^l\Delta_i}$.
\end{lemma}
{\em Proof}: 
Since $\bigwedge_{j=1}^l\Delta_j \leq_{\Sigma} \Delta_i$, for each $i \in [1, l]$, it holds that $D_{\bigwedge_{j=1}^l\Delta_j} \subseteq D_{\Delta_i}$. Thus, it holds that $I_{\Delta_i} \subseteq I_{\bigwedge_{j=1}^l\Delta_j}$ for each $i \in [1, l]$. We then have $\bigcup_{i \in [1, l]}I_{\Delta_i} \subseteq I_{\bigwedge_{i=1}^l\Delta_i}$.

On the other hand, let $(a, b) \in I_{\bigwedge_{i=1}^l\Delta_i}$ be any independent symbols for $\bigwedge_{i=1}^l\Delta_i$. We must have that $(a, b) \in \bigcup_{i \in [1, l]}I_{\Delta_i}$. Indeed, suppose on the contrary that $(a, b) \notin I_{\Delta_i}$, for each $i \in [1, l]$. Then, $\{a, b\} \sqsubseteq \Delta_i$ for each $i \in [1, l]$. It follows that $\{a, b\} \sqsubseteq \bigwedge_{i=1}^l \Delta_i$ and thus $(a, b) \notin I_{\bigwedge_{i=1}^l \Delta_i}$, which is a contradiction.  \hfill $\blacksquare$ \\

By Lemma~\ref{lemma: indered}, $\bigcup_{i \in [1, l]}I_{\Delta_i}=I_{\Delta}$ if and only if $I_{\bigwedge_{i=1}^l\Delta_i}=I_{\Delta}$ if and only if $D_{\bigwedge_{i=1}^l\Delta_i}=D_{\Delta}$. So, we can view the gap between Lemma~\ref{prop: nece} and Lemma~\ref{lemma: c} from a different perspective.  Lemma~\ref{lemma: c} requires the equality $\bigwedge_{i=1}^l\Delta_i=\Delta$, and Lemma~\ref{prop: nece} only requires the dependence relation (resp., the independence relation) induced by $\bigwedge_{i=1}^l\Delta_i$ to be equal to the dependence relation (resp., the independence relation) induced by $\Delta$.

\vspace{-4pt}
\subsection{Ordering on Candidate Reductions}
It is not difficult to check that the relation $\leq_{\Delta}$ is a partial ordering. Indeed, we shall now show that 
 $(S_{\Delta}, \leq_{\Delta})$ is a finite lattice. The proof follows closely that of Theorem~\ref{prop: finitelattice}.
\begin{theorem}
\label{propo: sfin}
$(S_{\Delta}, \leq_{\Delta})$ is a finite lattice.
\end{theorem}
{\em Proof}: 
It is not difficult to check that $(S_{\Delta}, \leq_{\Delta})$ is a partial order. In the partial order $(S_{\Delta}, \leq_{\Delta})$, 
\begin{enumerate}
\item 
the meet $P \wedge P'$ of any two elements $P, P'\in S_{\Delta}$ always exists, and it is defined to be the element $[{P \cup P'}]\in S_{\Delta}$. Recall that $P''=[{P \cup P'}]\in S_{\Delta}$ is obtained from $P \cup P'$ by keeping only the minimal distributions (in terms of the partial ordering $\leq_{\Sigma}$).
We now show the following.
\begin{enumerate}
\item $P'' \leq_{\Delta} P$ and $P'' \leq_{\Delta} P'$: this is straightforward from the definition of $P''$ and the definition of $\leq_{\Delta}$. Indeed, for any $\Delta_i \in P$, if $\Delta_i \in P''$, then $\Delta_i \leq_{\Sigma} \Delta_i$; otherwise, by the definition of $P''$, there must exist some $\Delta_j'' \in P''$ such that $\Delta_j'' <_{\Sigma} \Delta_i$ ($\Delta_i \notin P''$, it must have been removed due to the presence of some $\Delta_j'' \in P''$ where $\Delta_j'' <_{\Sigma} \Delta_i$). Thus, for any $\Delta_i \in P$, there exists some $\Delta'' \in P''$ such that $\Delta'' \leq_{\Sigma} \Delta_i$. That is,  $P'' \leq_{\Delta} P$. Similarly, we have $P'' \leq_{\Delta} P'$.
\item for any $P''' \in S_{\Delta}$, if $P''' \leq_{\Delta} P$ and $P''' \leq_{\Delta} P'$, then $P''' \leq_{\Delta} P''$: this is straightforward from the definition of $P''$ and the definition of $\leq_{\Delta}$. Indeed, for any $\Delta_i'' \in P''$, it holds that $\Delta_i'' \in P$ or $\Delta_i'' \in P'$, by the definition of $P''$. If $\Delta_i'' \in P$, then there exists some $\Delta_m''' \in P'''$ such that $\Delta_m''' \leq_{\Sigma} \Delta_i''$, since $P''' \leq_{\Delta} P$. If $\Delta_i'' \in P'$, then there exists $\Delta_k''' \in P'''$ such that $\Delta_k''' \leq_{\Sigma} \Delta_i''$, since $P''' \leq_{\Delta} P'$. That is, for any $\Delta_i'' \in P''$, there exists some $\Delta''' \in P'''$ such that $\Delta''' \leq_{\Sigma} \Delta_i''$. Thus,  $P''' \leq_{\Delta} P''$.
\end{enumerate}
\item the join $P \vee P'$ of any two elements $P, P'\in S_{\Delta}$ always exists, and it is defined to be the element $P''\in S_{\Delta}$ that is obtained from the set
\begin{center}
$\{\Delta'' \in \mathcal{M}(\Delta) \mid \exists \Delta_i \in P, \Delta_j' \in P', \Delta_i \vee \Delta_j' \leq_{\Sigma} \Delta''\}$
\end{center}
by keeping only those minimal distributions (in terms of the partial ordering $\leq_{\Sigma}$).
 We now show the following. 
\begin{enumerate}
\item $P \leq_{\Delta} P''$ and $P' \leq_{\Delta} P''$: this is straightforward from the definition of $P''$ and the definition of $\leq_{\Delta}$. Indeed, for any $\Delta'' \in P''$, we have $\Delta_i \vee \Delta_j' \leq_{\Sigma} \Delta''$ for some $\Delta_i \in P$, $\Delta_j' \in P'$. It then follows that $\Delta_i \leq_{\Sigma} \Delta''$ and $\Delta_j' \leq_{\Sigma} \Delta''$. Thus, $P \leq_{\Delta} P''$ and $P' \leq_{\Delta} P''$. 
\item for any $P''' \in S_{\Delta}$, if $P \leq_{\Delta} P'''$ and $P' \leq_{\Delta} P'''$, then $P'' \leq_{\Delta} P'''$: this is straightforward from the definition of $P''$ and the definition of $\leq_{\Delta}$. Indeed, for any $\Delta''' \in P'''$, there exists some $\Delta_i \in P$ and $\Delta_j' \in P'$ such that $\Delta_i \leq_{\Sigma} \Delta'''$ and $\Delta_j' \leq_{\Sigma} \Delta'''$. It follows that $\Delta_i \vee \Delta_j' \leq_{\Sigma} \Delta'''$. Since $P''' \in S_{\Delta}$, we have $\Delta''' \in \mathcal{M}(\Delta)$. Thus, either $\Delta''' \in P''$ or there exists some $\Delta'' \in P''$ such that $\Delta''<_{\Sigma} \Delta'''$. Thus, we conclude $P'' \leq_{\Delta} P'''$. \hfill $\blacksquare$\\
\end{enumerate}
\end{enumerate}

As an application, the greatest lower bound of $S_{\Delta}$ is equal to $\bigwedge_{P \in S_{\Delta}}P=[{\bigcup_{P \in S_{\Delta}} P}]=[\mathcal{M}(\Delta)]$, since we have $\{\Delta'\} \in S_{\Delta}$ for each\footnote{Clearly, each candidate reduction $\{\Delta'\} \in S_{\Delta}$ cannot be a reduction of $\Delta$ by definition. They are only padded in $S_{\Delta}$ for technical convenience. $\varnothing$ is the least upper bound of $S_{\Delta}$; in particular, it is possible that $P\vee P'=\varnothing$.} $\Delta' \in \mathcal{M}(\Delta)$.  Thus, with Lemma~\ref{prop:down}, we obtain the result that $\Delta$ has a reduction if and only if $[\mathcal{M}(\Delta)]$ is a reduction of $\Delta$ (cf. Theorem~\ref{thm: exiten}). And this is in fact the key result behind the conclusion that Technical Problem 2 can be reduced to Technical Problem 1.

\vspace{-10pt}
\section{Reduction Verification}
\label{section:RVd}
In this section, we address the problem of reduction verification. In Section~\ref{sub:CCE}, we shall first develop a technique for reduction refutation, by using candidate counter examples. In Section~\ref{sec: cr}, we will then focus on reduction validation. The reduction verification procedure is then given in Section~\ref{sub:RVP}.

\vspace{-8pt}
\subsection{Candidate Counter Examples for Reduction Refutation}
\label{sub:CCE}
Suppose a candidate reduction $P$ is not a reduction of the distribution $\Delta$. Then, we know that it is possible to construct a finite language counter example to refute $P$~\cite{LWSW17b}. That is, we can produce a finite language over $\Sigma$ such that it is decomposable with respect to each distribution in $P$ but not decomposable with respect to $\Delta$. In this section, we provide a template $L_{cand}$ of candidate counter examples for any distribution $\Delta$, which turns out to be quite effective in refuting candidate reductions that are not valid. We first need the next result, which has been shown in the proof of Proposition 29 in~\cite{LWSW17b}.
\begin{lemma}
If $P \in S_{\Delta}$ is not a reduction of $\Delta$, then every counter example for refuting $P$ is trace closed with respect to $I_{\Delta}$ but not decomposable with respect to $\Delta$.
\end{lemma}

Thus, as a candidate counter example $L_{cand}$ of $P$, we need to ensure that $L_{cand}$ is indeed trace-closed with respect to $I_{\Delta}$ but not decomposable with respect to $\Delta$. For any alphabet $\Sigma=\{\sigma_1, \sigma_2, \ldots, \sigma_{|\Sigma|}\}$ and any distribution $\Delta=(\Sigma_1, \Sigma_2, \ldots, \Sigma_n)$ of $\Sigma$, we generate the following set $L(\Sigma_j)$ of strings for each $j \in [1, n]$.
\vspace{-3pt}
\begin{center}
$L(\Sigma_j):=\sigma_1^{h_j(1)} \shuffle \sigma_2^{h_j(2)} \shuffle \ldots  \shuffle \sigma_{|\Sigma|}^{h_j(|\Sigma|)}$
\end{center}
\vspace{-3pt}
where, for each $i \in [1, |\Sigma|]$, $h_j(i):=1$ if $\sigma_i \in \Sigma_j$; $h_j(i):=j+1$ if $\sigma_i \notin \Sigma_j$. Now, let $L_{cand}=\bigcup_{i=1}^nL(\Sigma_j)$. We have the following.
\begin{lemma}
\label{lemma:deocm}
$L_{cand}$ is trace-closed with respect to $I_{\Delta}$ but not decomposable with respect to $\Delta$. 
\end{lemma}

{\em Proof}: 
By the construction, each $L(\Sigma_j)$ is trace-closed with respect to $I_{\Delta}$. Thus, $L_{cand}$ is also trace-closed with respect to $I_{\Delta}$. Let $s_0:=\sigma_1\sigma_2 \ldots \sigma_{|\Sigma|}$. By definition, $s_0 \notin L_{cand}$. It is clear that, for each $j \in [1, n]$, $P_j(s_0) \in P_j(L(\Sigma_j))$. Thus, 
\begin{center}
$s_0 \in \lVert_{j=1}^n P_j(s_0) \subseteq \lVert_{j=1}^n P_j(L(\Sigma_j)) \subseteq \lVert_{j=1}^n P_j(L_{cand})$
\end{center}
Thus, $s_0$ is a witness of the fact that $L_{cand}$ is not decomposable with respect to $\Delta$. 
\hfill $\blacksquare$ \\

We now use the next example to illustrate the construction of $L_{cand}$.
\begin{example}
Consider distribution $\Delta=(\Sigma_1, \Sigma_2, \Sigma_3, \Sigma_4)$ of $\Sigma=\{a, b, c, d\}$, where $\Sigma_1=\{a, b\}$, $\Sigma_2=\{b, c\}$, $\Sigma_3=\{c, d\}$ and $\Sigma_4=\{d, a\}$. Then, $L_{cand}=$
\begin{center}
$(a \shuffle b \shuffle c^2 \shuffle d^2) \cup (a^3\shuffle b \shuffle c\shuffle d^3) \cup (a^4\shuffle b^4\shuffle c\shuffle d) \cup (a\shuffle b^5\shuffle c^5\shuffle d)$
\end{center}
\end{example}

We will first present an application of the candidate counter example $L_{cand}$ to a distribution of ring-like structure of size 5.
\begin{example}
\label{examp: r5}
Consider distribution $\Delta=(\Sigma_1, \Sigma_2, \Sigma_3, \Sigma_4, \Sigma_5)$ of $\Sigma=\{a, b, c, d, e\}$, where $\Sigma_1=\{a, b\}$, $\Sigma_2=\{b, c\}$, $\Sigma_3=\{c, d\}$, $\Sigma_4=\{d, e\}$ and $\Sigma_5=\{e, a\}$. We can use $L_{cand}$ to show that $\Delta$ has no reduction. Indeed, we only need to show that $[\bot(\Delta)]=\{\Delta_1, \Delta_2, \Delta_3, \Delta_4, \Delta_5\}$ is not a reduction of $\Delta$, by Theorem~\ref{thm: exiten}, where 
\[   \left\{
\begin{array}{ll}
      \Delta_1=(\{a, b, c\}, \{c, d\}, \{d, e\}, \{e, a\}) \\
      \Delta_2=(\{b, c, d\}, \{a, b\}, \{d, e\}, \{e, a\}) \\
      \Delta_3=(\{c, d, e\}, \{a, b\}, \{b, c\}, \{e, a\}) \\
      \Delta_4=(\{d, e, a\}, \{a, b\}, \{b, c\}, \{c, d\}) \\
      \Delta_5=(\{a, b, e\}, \{b, c\}, \{c, d\}, \{d, e\}) 
\end{array} 
\right. \]
By Lemma~\ref{lemma:deocm}, we only need to show $L_{cand}$ is decomposable with respect to each $\Delta_i$ in $[\bot(\Delta)]$. In this case, $L_{cand}=$
\begin{center}
$(a\shuffle b\shuffle c^2\shuffle d^2\shuffle e^2) \cup (a^3 \shuffle b\shuffle c\shuffle d^3 \shuffle e^3) \cup (a^4 \shuffle b^4\shuffle c\shuffle d\shuffle e^4) \cup (a^5\shuffle b^5\shuffle c^5\shuffle d\shuffle e) \cup (a \shuffle b^6\shuffle c^6\shuffle d^6\shuffle e)$
\end{center}
It is not difficult\footnote{We shall provide an analysis later in Example~\ref{example:anally}.} to verify that $L_{cand}$ is indeed decomposable with respect to each distribution $\Delta_i$ in $[\bot(\Delta)]$. Thus, $[\bot(\Delta)]$ is not a reduction of $\Delta$, and $\Delta$ has no reduction. We note that, in this example, Proposition~\ref{theorem: char} fails to refute $[\bot(\Delta)]$. 
\end{example}

One may be tempted to think that the reason why the above distribution $\Delta$ has no reduction is because each symbol in $\Sigma$ is shared and the coupling between sub-alphabets in $\Delta$ is tight. This is not entirely correct, as can be seen in the next example.

\begin{example}
\label{examp: r4}
Consider the distribution $\Delta=(\Sigma_1, \Sigma_2, \Sigma_3, \Sigma_4)$ of $\Sigma=\{a, b, c, d\}$ of size 4, where $\Sigma_1=\{a, b\}$, $\Sigma_2=\{b, c\}$, $\Sigma_3=\{c, d\}$ and $\Sigma_4=\{d, a\}$. $\Delta$ is of ring-like structure of size 4. $\Delta$ does not have a reduction. This can be shown by using $L_{cand}$ on $[\bot(\Delta)]$ again, as in Example~\ref{examp: r5}. 
On the other hand, consider the distribution $\Delta'=(\Sigma_1, \Sigma_2, \Sigma_3, \Sigma_4, \Sigma_5)$ of $\Sigma=\{a, b, c, d\}$ of size 5, where $\Sigma_1=\{a, b\}$, $\Sigma_2=\{b, c\}$, $\Sigma_3=\{c, d\}$, $\Sigma_4=\{d, a\}$ and $\Sigma_5=\{a, c\}$. It is straightforward to see, by using the substitution-based proof technique (see~\cite{LWSW17b} or Section~\ref{sec: cr} for more details), that $\{\Delta_1, \Delta_2\}$ is a $(2, 3)$-reduction of $\Delta'$, where
\[   \left\{
\begin{array}{ll}
      \Delta_1=(\{a, b\}, \{b, c\}, \{a, c, d\}) \\
      \Delta_2=(\{a, b, c\}, \{c, d\}, \{d, a\}) \\
\end{array} 
\right. \]   
\end{example}
In the remaining of this section, we will study the following two related problems.
\begin{enumerate}
\item [A)] Given any candidate reduction $P$ of $\Delta$, which we would like to refute via $L_{cand}$, we only need to show $L_{cand}$ is decomposable with respect to each distribution $\Delta_i$ in $P$. This can be achieved by directly applying the definition of decomposability, that is, by checking $L_{cand}=L_{cand}^{\Delta_i}$. Since $L_{cand}$ has a very special structure, it is possible\footnote{We shall only show two reasoning rules (see Theorem~\ref{theo: proof} in this section and Theorem~\ref{theo: chac} in Appendix A, which is an improvement of Theorem~\ref{theo: proof}).} to discharge the verification obligations by using some efficient reasoning rules. This may potentially lead to a more manageable approach to verify the decomposability of $L_{cand}$, especially for the more difficult parameterized setup. B) will provide such an example, where we need to reason about the decomposability for an infinite number of $L_{cand}$'s, to show a family of distributions of ring-like structures do not have reductions. However, one could come up with several different reasoning rules (or, explanations) for the decomposability of $L_{cand}$. It is of much interest to develop an intuitive reasoning rule that is also powerful: If $L_{cand}$ is decomposable with respect to $\Delta_i$'s, then the reasoning rule shall often conclude it.
 Preferably, $L_{cand}$ is abstracted away in the analysis and we only need to work with sub-alphabets in $\Delta$ and $\Delta_i$'s, that is, explore structures of the distributions. Indeed, we believe that this can lead to a better understanding of the reason why a distribution does not have a reduction.
\item [B)] The distributions $\Delta$'s of ring-like structures provided in Examples~\ref{examp: r5} and~\ref{examp: r4} suggest that the distribution $\Delta=$
\begin{center}
$(\{a_0, a_1\}, \{a_1, a_2\}, \ldots, \{a_{n-2}, a_{n-1}\}, \{a_{n-1}, a_0\})$ 
\end{center}
of $\Sigma=\{a_0, a_1, \ldots, a_{n-1}\}$ may have no reduction, for any $n \geq 3$. It is of interest to know whether this is indeed true; and how to prove these parameterized results?
\end{enumerate}

To provide the intuition, we shall now analyze why $L_{cand}$ is decomposable with respect to each $\Delta_i$ in Example~\ref{examp: r5}. We only need to explain the decomposability of $L_{cand}$ with respect to distribution $\Delta_1$; for other distributions $\Delta_i$'s, the explanation is similar.

\begin{example}
\label{example:anally}
Consider the distributions $\Delta$ and $\Delta_1$ given in Example~\ref{examp: r5}. We recall here that $\Delta=(\Sigma_1, \Sigma_2, \Sigma_3, \Sigma_4, \Sigma_5)$, where $\Sigma_1=\{a, b\}$, $\Sigma_2=\{b, c\}$, $\Sigma_3=\{c, d\}$, $\Sigma_4=\{d, e\}$ and $\Sigma_5=\{e, a\}$. And, $\Delta_1=(\{a, b, c\}, \{c, d\}, \{d, e\}, \{e, a\})$.  We have $L_{cand}=L(\Sigma_1) \cup L(\Sigma_2) \cup L(\Sigma_3) \cup L(\Sigma_4) \cup L(\Sigma_5)=$
\begin{center}
$(a\shuffle b\shuffle c^2\shuffle d^2\shuffle e^2) \cup (a^3 \shuffle b\shuffle c\shuffle d^3 \shuffle e^3) \cup (a^4 \shuffle b^4\shuffle c\shuffle d\shuffle e^4) \cup (a^5\shuffle b^5\shuffle c^5\shuffle d\shuffle e) \cup (a \shuffle b^6\shuffle c^6\shuffle d^6\shuffle e)$
\end{center}
To verify the decomposability of $L_{cand}$ with respect to $\Delta_1$, we only need to check whether $L_{cand}^{\Delta_1} \subseteq L_{cand}$. The decomposition closure $L_{cand}^{\Delta_1}$ of $L_{cand}$ is by definition equal to
\begin{center}
$\bigcup_{Dom} P_{\{a, b, c\}}(s_1) \lVert P_{\{c, d\}}(s_2) \lVert P_{\{d, e\}}(s_3) \lVert P_{\{e, a\}}(s_4)$ 
\end{center}
where $Dom:=\bigwedge_{i \in [1, 4]}s_i \in L_{cand}$ and we have used $P_{\Sigma'}$ to denote the projection from $\Sigma^*$ to $\Sigma'^*$ for $\Sigma' \subseteq \Sigma$. We only need to show that, for any $(s_1, s_2, s_3, s_4)$ that satisfies $Dom$, if
\begin{center}
$P_{\{a, b, c\}}(s_1) \lVert P_{\{c, d\}}(s_2) \lVert P_{\{d, e\}}(s_3) \lVert P_{\{e, a\}}(s_4) \neq \varnothing$,
\end{center}
then we must have
\begin{center}
 $P_{\{a, b, c\}}(s_1) \lVert P_{\{c, d\}}(s_2) \lVert P_{\{d, e\}}(s_3) \lVert P_{\{e, a\}}(s_4) \subseteq L_{cand}$.  
\end{center}

We shall perform case analysis on $s_1 \in L_{cand}$; and there are five cases that correspond to $s_1 \in L(\Sigma_j)$, where $j \in [1, 5]$. We shall only analyze the cases $s_1 \in L(\Sigma_1)$ and $s_1 \in L(\Sigma_2)$; and the rest cases can be analyzed in a similar way.

$(s_1 \in L(\Sigma_1))$: Suppose we select $s_1 \in L(\Sigma_1)=a\shuffle b\shuffle c^2\shuffle d^2\shuffle e^2$, then we have $P_{\{a, b, c\}}(s_1) \in a\shuffle b\shuffle c^2$. We then need to select $s_2, s_3, s_4$ so that 
\begin{center}
$P_{\{a, b, c\}}(s_1) \lVert P_{\{c, d\}}(s_2) \lVert P_{\{d, e\}}(s_3) \lVert P_{\{e, a\}}(s_4) \neq \varnothing$.
\end{center}
This determines that the only choice for $s_2$ is to choose $s_2 \in L(\Sigma_1)=a\shuffle b\shuffle c^2\shuffle d^2\shuffle e^2$, since any other choice of $s_2$ does not contain exactly two occurrences of the symbol $c$ that matches $c^2$ in $P_{\{a, b, c\}}(s_1)$. Then, $P_{\{c, d\}}(s_2) \in c^2\shuffle d^2$. Then, we need to select $s_3, s_4$ so that 
\begin{center}
$P_{\{a, b, c\}}(s_1) \lVert P_{\{c, d\}}(s_2) \lVert P_{\{d, e\}}(s_3) \lVert P_{\{e, a\}}(s_4) \neq \varnothing$.
\end{center}
$P_{\{c, d\}}(s_2) \in c^2\shuffle d^2$ determines that the only choice for $s_3$ is to choose $s_3 \in L(\Sigma_1)=a\shuffle b\shuffle c^2\shuffle d^2\shuffle e^2$ to match the number of occurrences of symbol $d$. We here observe that there is even no need to consider $s_4$ since $P_{\{a, b, c\}}(s_1)$, $P_{\{c, d\}}(s_2)$ and $P_{\{d, e\}}(s_3)$ have already covered each symbol $a, b, c, d, e$ in $\Sigma$. That is, 
\begin{center}
$P_{\{a, b, c\}}(s_1) \lVert P_{\{c, d\}}(s_2) \lVert P_{\{d, e\}}(s_3) \lVert P_{\{e, a\}}(s_4) \subseteq P_{\{a, b, c\}}(s_1) \lVert P_{\{c, d\}}(s_2) \lVert P_{\{d, e\}}(s_3) \subseteq (a\shuffle b\shuffle c^2) \lVert (c^2\shuffle d^2) \lVert (d^2 \shuffle e^2) \subseteq a\shuffle b\shuffle c^2\shuffle d^2\shuffle e^2=L(\Sigma_1) \subseteq L_{cand}$ \qed
\end{center}

That is, for any $s_1 \in L(\Sigma_1)$, for any choice of $s_2, s_3, s_4 \in L_{cand}$ such that \begin{center}
$P_{\{a, b, c\}}(s_1) \lVert P_{\{c, d\}}(s_2) \lVert P_{\{d, e\}}(s_3) \lVert P_{\{e, a\}}(s_4) \neq \varnothing$,
\end{center}
we always have
\begin{center}
 $P_{\{a, b, c\}}(s_1) \lVert P_{\{c, d\}}(s_2) \lVert P_{\{d, e\}}(s_3) \lVert P_{\{e, a\}}(s_4) \subseteq L_{cand}$.  
\end{center} 

$(s_1 \in L(\Sigma_2))$: Suppose we select $s_1 \in L(\Sigma_2)=a^3 \shuffle b\shuffle c\shuffle d^3 \shuffle e^3$, then we have $P_{\{a, b, c\}}(s_1) \in a^3 \shuffle b\shuffle c$. Now, $P_{\{a, b, c\}}(s_1) \in a^3 \shuffle b\shuffle c$ determines that the only choice for $s_4$ is to set $s_4 \in L(\Sigma_2)=a^3 \shuffle b\shuffle c\shuffle d^3 \shuffle e^3$, to match the number of occurrences of symbol $a$, and thus $P_{\{e, a\}}(s_4)=a^3\shuffle e^3$; and $P_{\{e, a\}}(s_4)=a^3\shuffle e^3$ determines the only choice for $s_3$ to be $s_3 \in L(\Sigma_2)=a^3 \shuffle b\shuffle c\shuffle d^3 \shuffle e^3$, to match the number of occurrences of symbol $e$, and thus $P_{\{d, e\}}(s_3)=d^3\shuffle e^3$. We note that $P_{\{a, b, c\}}(s_1)$, $P_{\{e, a\}}(s_4)$ and $P_{\{d, e\}}(s_3)$ have already covered each symbol $a, b, c, d, e$ in $\Sigma$. Thus, 
\begin{center}
$P_{\{a, b, c\}}(s_1) \lVert P_{\{c, d\}}(s_2) \lVert P_{\{d, e\}}(s_3) \lVert P_{\{e, a\}}(s_4) \subseteq P_{\{a, b, c\}}(s_1) \lVert P_{\{d, e\}}(s_3) \lVert P_{\{e, a\}}(s_4) \subseteq (a^3 \shuffle b\shuffle c) \lVert (d^3\shuffle e^3) \lVert (a^3\shuffle e^3) \subseteq a^3 \shuffle b\shuffle c\shuffle d^3 \shuffle e^3=L(\Sigma_2) \subseteq L_{cand}$ \qed
\end{center}

That is, for any $s_1 \in L(\Sigma_2)$, for any choice of $s_2, s_3, s_4 \in L_{cand}$ such that \begin{center}
$P_{\{a, b, c\}}(s_1) \lVert P_{\{c, d\}}(s_2) \lVert P_{\{d, e\}}(s_3) \lVert P_{\{e, a\}}(s_4) \neq \varnothing$,
\end{center}
we always have
\begin{center}
 $P_{\{a, b, c\}}(s_1) \lVert P_{\{c, d\}}(s_2) \lVert P_{\{d, e\}}(s_3) \lVert P_{\{e, a\}}(s_4) \subseteq L_{cand}$.  
\end{center} 

Thus, we conclude that $L_{cand}$ is decomposable with respect to $\Delta_1$, after the analysis of the above five cases.
\end{example}
It is beneficial to look at the above reasoning from a more abstract point of view, which is conducted below.
\begin{example}
\label{exa: analys}
The analysis on the five cases $s_1 \in L(\Sigma_j)$'s can be briefly summarized as follows. $P_{\{a, b, c\}}(s_1)$ determines (the numbers of occurrences of) symbols $a, b, c$. In order to cover all the symbols $a, b, c, d, e$, we only need to further determine $d$ and $e$. We shall only explain how $d$ can be determined. The case for $e$ is quite similar. 

i) For any string $s_1$ in $L(\Sigma_1), L(\Sigma_4)$ or $L(\Sigma_5)$, the number of occurrences of $c$ is distinctive\footnote{The number of occurrences of $c$ is not distinctive for $s_1$ in $L(\Sigma_2)$, since strings in $L(\Sigma_2)$ and $L(\Sigma_3)$ have the same number of occurrences of $c$. The same conclusion holds for $L(\Sigma_3)$. For simplicity, we also say $c$ is a distinctive symbol for $s_1 \in L(\Sigma_j)$, $j=1, 4, 5$. Clearly, $c$ is a distinctive symbol for $s_1 \in L(\Sigma_j)$ iff $c \notin \Sigma_j$.} and different from $1$ by the construction, since $c \notin \Sigma_1, \Sigma_4, \Sigma_5$. In these cases, $P_{\{a, b, c\}}(s_1)$ then determines the symbol $d$ through $P_{\{c, d\}}(s_2)$ by matching the symbol $c$. The case $s_1 \in L(\Sigma_1)$ discussed previously serves as an illustration.  

Thus, in Case i), $d$ is determined via $(c, d) \in D_{\Delta_1}$, when $c$ is a distinctive symbol for $s_1$, i.e., when $s_1$ is in $L(\Sigma_1), L(\Sigma_4)$ or $L(\Sigma_5)$. 

ii) On the other hand, for any $s_1$ in $L(\Sigma_2)$ or $L(\Sigma_3)$, the number of occurrences of $c$ is not distinctive, since $c \in \Sigma_2, \Sigma_3$. Thus, in these cases, $P_{\{a, b, c\}}(s_1)$ cannot be used to determine symbol $d$  by matching $c$. Instead, in these cases, $P_{\{a, b, c\}}(s_1)$ can be used to first determine symbol $e$, through $P_{\{e, a\}}(s_4)$, by matching $a$ (after the matching, $s_1$ and $s_4$ belong to the same $L(\Sigma_j)$, $j=2, 3$). Then, $P_{\{e, a\}}(s_4)$ can be used to determine the symbol $d$, through $P_{\{d, e\}}(s_3)$, by matching $e$. The chain of the matching can be performed since neither $a$ nor $e$ belongs to $\Sigma_2$, $\Sigma_3$ (the numbers of occurrences of $a$ and $e$ in $L(\Sigma_2)$ and $L(\Sigma_3)$ are thus distinctive, which can be used for matching). The case $s_1 \in L(\Sigma_2)$ discussed before is an illustration. 

Thus, in Case ii), $d$ is determined via the chain $(a, e)(e, d)$, where $(a, e)$ and $(e, d)$ are elements of $D_{\Delta_1}$. For the purpose of matching, both $a$ and $e$ need to be distinctive. To make $a$ distinctive, we require $s_1$ to be in $L(\Sigma_2), L(\Sigma_3)$ or $L(\Sigma_4)$. To make $e$ distinctive, we require $s_1$ to belong to $L(\Sigma_1), L(\Sigma_2)$ or $L(\Sigma_3)$. Thus, $d$ is determined via the chain $(a, e)(e, d)$ when $s_1$ is in $L(\Sigma_2)$ or $L(\Sigma_3)$.

Thus, for any choice of $s_1$, symbol $d$ is determined. A similar analysis shows that, for any choice of $s_1$, $e$ is determined. That is, for any choice of $s_1$, we can cover each symbol $a, b, c, d, e$ in $\Sigma$. Thus, we can now conclude that $L_{cand}$ is decomposable with respect to $\Delta_1$. The analysis result is summarized in the following table. The second column shows how $d$ is determined for $s_1 \in L(\Sigma_j)$, $j \in [1, 5]$; and the third column shows how $e$ is determined.

\begin{center}
\begin{tabular}{ |c|c|c|c| } 
\hline
a, b, c & d & e \\
\hline
$s_1 \in L(\Sigma_1)$ & (c, d) & (c, d)(d, e) \\ 
\hline
$s_1 \in L(\Sigma_2)$ & (a, e)(e, d) & (a, e) \\ 
\hline
$s_1 \in L(\Sigma_3)$ & (a, e)(e, d) & (a, e) \\ 
\hline
$s_1 \in L(\Sigma_4)$ & (c, d) & (a, e) \\ 
\hline
$s_1 \in L(\Sigma_5)$ & (c, d) & (c, d)(d, e) \\ 
\hline
\end{tabular}
\end{center}
Such a table can be viewed as a proof of the decomposability of $L_{cand}$ with respect to $\Delta_1$. 
Thus, our goal is to synthesize such a table, if possible, to discharge the verification obligation. We shall note that essentially we only need to produce certain paths (satisfying certain constraints) in the graph $(\Sigma, D_{\Delta_1})$ to synthesize such a table (c. f. second and third columns). This motivates us to develop a dependence graph based technique for verifying the decomposability of $L_{cand}$ with respect to the distributions in $\mathcal{M}(\Delta)$. The reason we analyze the choice of $s_1$ is that the corresponding sub-alphabet $\{a, b, c\}$ in $\Delta_1$ does not appear in $\Delta$. Thus, there must exist some distinctive symbols in $\{a, b, c\}$ that we can use for matching, for each choice of $s_1$.
\end{example}

We now present the dependence graph based technique for verifying the decomposability of $L_{cand}$, which shall reflect the analysis carried out in Example~\ref{exa: analys} from a more abstract point of view. To that end, we shall introduce some technical notation (see Appendix A for an improvement).

In the following, let $\Delta=(\Sigma_1, \Sigma_2, \ldots, \Sigma_n)$ be any distribution of $\Sigma$ and let $\Delta'=(\Sigma_1', \Sigma_2', \ldots, \Sigma_k')$ be any distribution of $\Sigma$ such that $\Delta \leq_{\Sigma} \Delta'$. Then, $D_{\Delta'}$ is the dependence relation induced by $\Delta'$, which can be visualized by using the graph $(\Sigma, D_{\Delta'})$. In addition, we also define a map $\mathcal{N}: \Sigma \rightarrow 2^{[1, n]}$ such that each\footnote{We here shall note that the graph $(\Sigma, D_{\Delta'})$ is defined based on $\Delta'$ and $\mathcal{N}$ is defined based on $\Delta$. $\mathcal{N}(\sigma)$ records the set of indexes $j$'s for which $\sigma$ is a distinctive symbol for $s \in L(\Sigma_j)$.} $\mathcal{N}(\sigma):=\{i \in [1, n] \mid \sigma \notin \Sigma_i\}$ is used to record the set of indexes of those sub-alphabets in $\Delta$ to which $\sigma$ does not belong. Given any symbol $\sigma \in \Sigma'$ and any symbol $\sigma' \in \Sigma-\Sigma'$, the set of all simple paths from $\sigma$ to $\sigma'$ in graph $(\Sigma, D_{\Delta'})$ is denoted by $SP_{\Delta', \Sigma'}(\sigma, \sigma')$. And, for each path $p=\sigma_1\sigma_2\ldots \sigma_m \in SP_{\Delta', \Sigma'}(\sigma, \sigma')$, let $\overline{Cr}_{\Delta'}(p):=\bigcap_{i=1}^{m-1}\mathcal{N}(\sigma_i)$. Then, set $Cr_{\Delta'}(\sigma, \sigma'):=\bigcup_{p \in SP_{\Delta', \Sigma'}(\sigma, \sigma')}\overline{Cr}_{\Delta'}(p)$. Finally, we let $Cr_{\Delta'}(\Sigma', \sigma'):=\bigcup_{\sigma \in \Sigma'}Cr_{\Delta'}(\sigma, \sigma')$. Then, we have the following method for testing the decomposability of $L_{cand}$.
\begin{theorem}
\label{theo: proof}
Let $\Delta'=(\Sigma_1', \Sigma_2', \ldots, \Sigma_k') \in \mathcal{M}(\Delta)$. $L_{cand}$ is decomposable with respect to $\Delta'$ if there exists a sub-alphabet $\Sigma_i' \in \Delta'-\Delta$ such that for each $\sigma' \in \Sigma-\Sigma_i'$, $Cr_{\Delta'}(\Sigma_i', \sigma')=[1, n]$, where $\Delta=(\Sigma_1, \Sigma_2, \ldots, \Sigma_n)$.
\end{theorem}

{\em Proof}: Suppose the sub-alphabet $\Sigma_i' \in \Delta'-\Delta$ satisfies the property that for each $\sigma' \in \Sigma-\Sigma_i'$, $Cr_{\Delta'}(\Sigma_i', \sigma')=[1, n]$, where $i$ is some integer in $[1, k]$. To show that $L_{cand}$ is decomposable with respect to $\Delta'$, we only need to show that for any $s_1, s_2, \ldots, s_i, \ldots s_k \in L_{cand}$, where $L_{cand}=\bigcup_{j=1}^n L(\Sigma_j)$, if
\begin{center}
$P_{\Sigma_1'}(s_1) \lVert P_{\Sigma_2'}(s_2) \lVert \ldots \lVert P_{\Sigma_i'}(s_i) \lVert \ldots \lVert P_{\Sigma_{k}'}(s_{k}) \neq \varnothing$,
\end{center}
then we must have
\begin{center}
$P_{\Sigma_1'}(s_1) \lVert P_{\Sigma_2'}(s_2) \lVert \ldots \lVert P_{\Sigma_i'}(s_i) \lVert \ldots \lVert P_{\Sigma_{k}'}(s_{k}) \subseteq L_{cand}$.
\end{center}

We shall perform case analysis on $s_i \in L_{cand}$, for which the sub-alphabet $\Sigma_i'$ belongs to $\Delta'-\Delta$; and there are $n$ cases that correspond to $s_i \in L(\Sigma_j)$, where $j \in [1, n]$. We consider an arbitrary fixed integer $j \in [1, n]$ and then analyze the case\footnote{For an illustration, Example~\ref{example:anally} analyzes the choice of $s_1 \in L_{cand}$ with $\Sigma_1'=\{a, b, c\} \in \Delta'-\Delta$.} $s_i \in L(\Sigma_j)$. 

For any $s_i \in L(\Sigma_j)$, the symbols in $\Sigma_i'$ have already been determined (via $P_{\Sigma_i'}(s_i)$). We shall first show that each symbol in $\Sigma-\Sigma_i'$ is also determined. Let $\sigma'$ be any symbol in $\Sigma-\Sigma_i'$. We conclude that $j \in [1, n]=Cr_{\Delta'}(\Sigma_i', \sigma')$. By the definition of $Cr_{\Delta'}(\Sigma_i', \sigma')$, we know that there exists some symbol $\sigma \in \Sigma_i'$ and some (simple) path $p \in SP_{\Delta', \Sigma_i'}(\sigma, \sigma')$ such that $j \in \overline{Cr}_{\Delta'}(p)$; now suppose $p=\sigma_1\sigma_2\ldots \sigma_m$, where $\sigma_1=\sigma$ and $\sigma_m=\sigma'$, then, $j \in \mathcal{N}(\sigma_h)$ for each $h \in [1, m-1]$. It then follows that $\sigma_h \notin \Sigma_j$, for each $h \in [1, m-1]$.  So, we know that $\sigma=\sigma_1 \notin \Sigma_j$ is a distinctive symbol for $s_i$ in $L(\Sigma_j)$; thus, it can be used for matching and determining $\sigma_2$, since $(\sigma, \sigma_2) \in D_{\Delta'}$. If $m=2$, then $\sigma'=\sigma_m$ has been determined; otherwise, $\sigma_2 \notin \Sigma_j$ is a distinctive symbol for any (matched) string in $L(\Sigma_j)$ and it can be used to further determine $\sigma_3$, since we have $(\sigma_2, \sigma_3) \in D_{\Delta'}$. We only need to continue the same analysis, and it then follows that $s_i \in L(\Sigma_j)$ determines the symbol $\sigma_m=\sigma'$. Now, since $\sigma'$ is an arbitrary symbol in $\Sigma-\Sigma_i'$, we conclude that $s_i \in L(\Sigma_j)$ determines each symbol in $\Sigma-\Sigma_i$; thus, for any choice of $s_i$ in $L(\Sigma_j)$, all the symbols in $\Sigma$ are covered. 

Since $j \in [1, n]$ is arbitrary, we then conclude that, for any choice of $s_i \in L_{cand}$, all the symbols in $\Sigma$ are covered; that is, $L_{cand}$ is decomposable with respect to $\Delta'$.
\hfill $\blacksquare$ \\

It is possible to avoid unnecessary computation involved in the definition of $Cr_{\Delta'}(\Sigma', \sigma')$ as follows. The set of boundary symbols in $\Sigma' \subseteq \Sigma$ (with respect to the graph $(\Sigma, D_{\Delta'})$) is defined\footnote{That is, a symbol $\sigma$ in $\Sigma'$ is a boundary symbol if it is adjacent to some symbol $\sigma' \in \Sigma-\Sigma'$ in $(\Sigma, D_{\Delta'})$.} to be 
$B_{\Delta'}(\Sigma'):=\{\sigma \in \Sigma' \mid \exists \sigma' \in \Sigma-\Sigma', (\sigma, \sigma') \in D_{\Delta'}\}$. 
Then, for any boundary symbol $\sigma \in B_{\Delta'}(\Sigma') \subseteq \Sigma'$ and any symbol $\sigma' \in \Sigma-\Sigma'$, $SP_{\Delta', \Sigma'}^R(\sigma, \sigma')$ shall be used to denote the set of all simple paths from $\sigma$ to $\sigma'$ in graph $(\Sigma, D_{\Delta'})$ that do not pass through any symbol within $\Sigma'$ (except for $\sigma$). Then, we shall let $Cr_{\Delta'}^R(\sigma, \sigma'):=\bigcup_{p \in SP_{\Delta', \Sigma'}^R(\sigma, \sigma')}\overline{Cr}_{\Delta'}(p)$. Finally, we define $Cr_{\Delta'}^R(\Sigma', \sigma'):=\bigcup_{\sigma \in B_{\Delta'}(\Sigma')}Cr_{\Delta'}^R(\sigma, \sigma')$. By definition, it is clear that $SP_{\Delta', \Sigma'}^R(\sigma, \sigma') \subseteq SP_{\Delta', \Sigma'}(\sigma, \sigma')$ and $B_{\Delta'}(\Sigma') \subseteq \Sigma'$. Thus, we conclude that $Cr_{\Delta'}^R(\sigma, \sigma') \subseteq  Cr_{\Delta'}(\sigma, \sigma')$ and $Cr_{\Delta'}^R(\Sigma', \sigma') \subseteq Cr_{\Delta'}(\Sigma', \sigma')$. 

For any symbol $\sigma \in \Sigma'$ and any path $p \in SP_{\Delta', \Sigma'}(\sigma, \sigma')$, it is clear that $p$ must pass through some boundary symbols, since $\sigma' \in \Sigma-\Sigma'$. Let $\sigma'' \in B_{\Delta'}(\Sigma')$ be the last boundary symbol that is passed through by $p$. Now, let $p'=p[\sigma'' \ldots]$ denote the segment of the path $p$ that starts from $\sigma''$. It is clear that $p' \in SP_{\Delta', \Sigma'}^R(\sigma'', \sigma')$ and $\overline{Cr}_{\Delta'}(p) \subseteq \overline{Cr}_{\Delta'}(p')$. Thus, $Cr_{\Delta'}(\Sigma', \sigma') \subseteq Cr_{\Delta'}^R(\Sigma', \sigma')$ holds and we can then conclude that $Cr_{\Delta'}(\Sigma', \sigma')=Cr_{\Delta'}^R(\Sigma', \sigma')=\bigcup_{\sigma \in B_{\Delta'}(\Sigma')}Cr_{\Delta'}^R(\sigma, \sigma')$, where $Cr_{\Delta'}^R(\sigma, \sigma'):=\bigcup_{p \in SP_{\Delta', \Sigma'}^R(\sigma, \sigma')}\overline{Cr}_{\Delta'}(p)$. We have the following remark.

\begin{figure}[t]
\centering
\hspace*{-1mm}
\includegraphics[width=2.0in, height=1.0in]{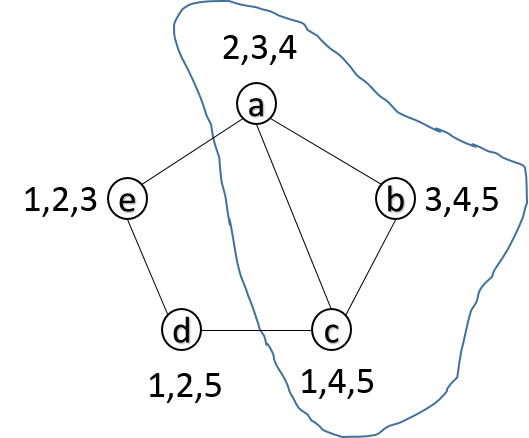} 
\captionsetup{justification=centering}
\caption{graph $(\Sigma, D_{\Delta_1})$, where self-loops are ignored}
\label{fig:dep}
\end{figure}
\begin{remark}
Intuitively, $j \in Cr_{\Delta'}(\Sigma_i', \sigma')$ has the meaning that, for any choice of $s_i \in L(\Sigma_j)$, symbol $\sigma'$ is determined. Thus, $Cr_{\Delta'}(\Sigma_i', \sigma')=[1, n]$ means that, for any $s_i \in L_{cand}$, symbol $\sigma'$ is determined. This is also reflected in the proof of Theorem~\ref{theo: proof}.
\end{remark}
We shall now illustrate the use of Theorem~\ref{theo: proof}.
\begin{example}
\label{ex:ills}
Let us now revisit Example~\ref{examp: r5}.
Here, we explain why $L_{cand}$ is decomposable with respect to $\Delta_1$. The graph $(\Sigma, D_{\Delta_1})$ is shown in Fig.~\ref{fig:dep}, where the map $\mathcal{N}$ is also drawn. The only element in $\Delta_1-\Delta$ is the sub-alphabet $\{a, b, c\}$. By Theorem~\ref{theo: proof}, we only need to show that both $Cr_{\Delta_1}(\{a, b, c\}, d)$ and $Cr_{\Delta_1}(\{a, b, c\}, e)$ are equal to $[1, 5]$.

1) The set $B_{\Delta_1}(\{a, b, c\})$ of boundary symbols in $\{a, b, c\}$ is $\{a, c\}$. Thus, $Cr_{\Delta_1}(\{a, b, c\}, d)=Cr_{\Delta_1}^R(a, d) \cup Cr_{\Delta_1}^R(c, d)$. Now, the set $SP_{\Delta_1, \{a, b, c\}}^R(a, d)$ of all simple paths from $a$ to $d$ in the graph $(\Sigma, D_{\Delta_1})$ that do not pass through any symbol within $\{a, b, c\}$ (except for $a$) is the singleton $\{aed\}$. Thus, $Cr_{\Delta_1}^R(a, d)=\overline{Cr}_{\Delta_1}(aed)=\mathcal{N}(a) \cap \mathcal{N}(e)=\{2, 3\}$. Similarly, $Cr_{\Delta_1}^R(c, d)=\overline{Cr}_{\Delta_1}(cd)=\mathcal{N}(c)=\{1, 4, 5\}$. Thus, we have $Cr_{\Delta_1}(\{a, b, c\}, d)=[1, 5]$.

2) Similarly, $Cr_{\Delta_1}(\{a, b, c\}, e)=Cr_{\Delta_1}^R(a, e) \cup Cr_{\Delta_1}^R(c, e)$. It is easy to check that $Cr_{\Delta_1}^R(a, e)=\overline{Cr}_{\Delta_1}(ae)=\mathcal{N}(a)=\{2, 3, 4\}$; and $Cr_{\Delta_1}^R(c, e)=\overline{Cr}_{\Delta_1}(cde)=\mathcal{N}(c) \cap \mathcal{N}(d)=\{1, 5\}$. Thus, we have $Cr_{\Delta_1}(\{a, b, c\}, d)=[1, 5]$.

The decomposability of $L_{cand}$ with respect to $\Delta_2, \Delta_3, \Delta_4$ and $\Delta_5$ can also be easily explained by using Theorem~\ref{theo: proof}. 

\end{example}

For any distribution $\Delta' \in \mathcal{M}(\Delta)$, $D_{\Delta} \subseteq D_{\Delta'}$ holds. Thus, $(\Sigma, D_{\Delta'})$ has strictly more edges than $(\Sigma, D_{\Delta})$ does. Instead of computing $Cr_{\Delta'}(\Sigma_i', \sigma')$, we can compute\footnote{That is, we work on the graph $(\Sigma, D_{\Delta})$, where $\mathcal{N}$ is defined as before.} $Cr_{\Delta}(\Sigma_i', \sigma')$, which is also well-defined for any sub-alphabet $\Sigma_i'$ in $\Delta'$ and any $\sigma \in \Sigma-\Sigma_i'$; furthermore\footnote{This is due to $SP_{\Delta, \Sigma_i'}(\sigma, \sigma') \subseteq SP_{\Delta', \Sigma_i'}(\sigma, \sigma')$ for any $\sigma \in \Sigma_i'$ and $\overline{Cr}_{\Delta'}(p)=\overline{Cr}_{\Delta}(p)$ for any $p \in SP_{\Delta, \Sigma_i'}(\sigma, \sigma')$.}, $Cr_{\Delta}(\Sigma_i', \sigma') \subseteq Cr_{\Delta'}(\Sigma_i', \sigma')$. Thus, we immediately have the following, which allows us to work on the graph $(\Sigma, D_{\Delta})$ uniformly for each $\Delta' \in \mathcal{M}(\Delta)$.
\begin{corollary}
\label{coro: simpli}
Let $\Delta'=(\Sigma_1', \Sigma_2', \ldots, \Sigma_k') \in \mathcal{M}(\Delta)$. $L_{cand}$ is decomposable with respect to $\Delta'$ if there exists a sub-alphabet $\Sigma_i' \in \Delta'-\Delta$ such that for each $\sigma' \in \Sigma-\Sigma_i'$, $Cr_{\Delta}(\Sigma_i', \sigma')=[1, n]$, where $\Delta=(\Sigma_1, \Sigma_2, \ldots, \Sigma_n)$.
\end{corollary}

{\em Proof}: This immediately follows from Theorem~\ref{theo: proof} and the fact that $Cr_{\Delta}(\Sigma_i', \sigma') \subseteq Cr_{\Delta'}(\Sigma_i', \sigma')$ for each $\Sigma_i' \in \Delta'$ and each $\sigma' \in \Sigma-\Sigma_i'$. \hfill $\blacksquare$ \\

To prove $\Delta$ has no reduction, we have the following result.
\begin{corollary}
\label{coro:suffno}
$\Delta$ does not have a reduction if for any two different sub-alphabets $\Sigma_i, \Sigma_j \in \Delta$ such that $\Sigma_i \cup \Sigma_j \neq \Sigma$, it holds that for each $\sigma' \in \Sigma-(\Sigma_i \cup \Sigma_j)$, $Cr_{\Delta}(\Sigma_i \cup \Sigma_j, \sigma')=[1, n]$, where $\Delta=(\Sigma_1, \Sigma_2, \ldots, \Sigma_n)$.
\end{corollary}

{\em Proof}: We have that $\Delta$ does not have a reduction iff $[\bot(\Delta)]$ is not a reduction of $\Delta$ iff $\bot(\Delta)$ is not a reduction of $\Delta$ iff $L_{cand}$ is decomposable with respect to each distribution in $\bot(\Delta)$. The only sub-alphabet in $\Delta'-\Delta$ for each $\Delta' \in \bot(\Delta)$ is the union of some sub-alphabets $\Sigma_i$, $\Sigma_j$ in $\Delta$ such that $\Sigma_i \cup \Sigma_j \neq \Sigma$. The statement then immediately follows from Corollary~\ref{coro: simpli}.  \hfill $\blacksquare$ \\

Corollary~\ref{coro:suffno} can be used to show the following ``parameterized" result.
\begin{theorem}
\label{them: non-exi}
For any $n \geq 3$, the distribution 
\vspace{-3pt}
\begin{center}
$\Delta=(\{a_0, a_1\}, \{a_1, a_2\}, \ldots, \{a_{n-2}, a_{n-1}\}, \{a_{n-1}, a_0\})$ 
\end{center}
\vspace{-3pt}
of $\Sigma=\{a_0, a_1, \ldots, a_{n-1}\}$ does not have a reduction. 
\end{theorem}
{\em Proof}: $\Delta$ has no reduction when $n=3$, since $\mathcal{M}(\Delta)=\varnothing$. And, from Example~\ref{examp: r4} and Example~\ref{examp: r5}, we know that $\Delta$ does not have a reduction when $n=4$ or $n=5$. Thus, we assume $n \geq 6$. Let $\Sigma_i:=\{a_i, a_{i+1}\}$ for each $i \in [0, n-1]$, where the addition $i+1$ is modulo $n$. 
Let $i_1<i_2$ be any two indexes in $[0, n-1]$. Clearly, $\Sigma_{i_1} \cup \Sigma_{i_2} \neq \Sigma$. We only need to show that for each $\sigma' \in \Sigma-(\Sigma_{i_1} \cup \Sigma_{i_2})$, $Cr_{\Delta}(\Sigma_{i_1} \cup \Sigma_{i_2}, \sigma')=[1, n]$.

\begin{figure}[t]
\centering
\hspace*{-1mm}
\includegraphics[width=2.6in, height=1.0in]{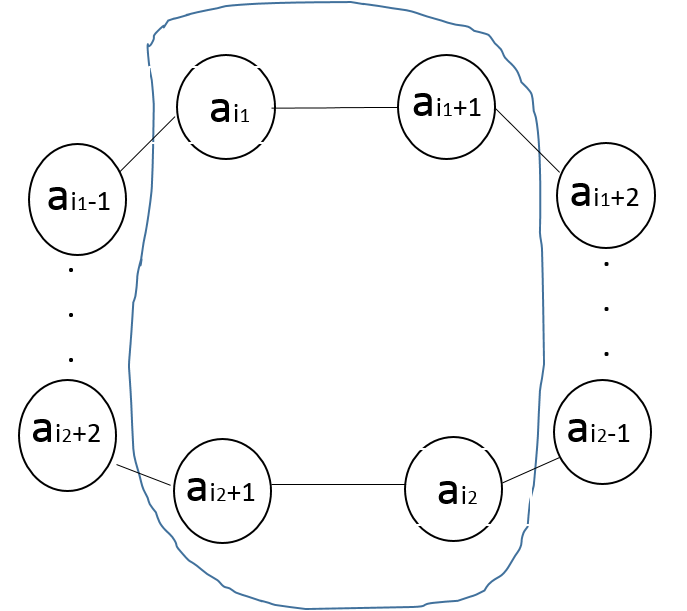} 
\captionsetup{justification=centering}
\caption{graph $(\Sigma, D_{\Delta})$}
\label{fig:ring}
\end{figure}

We first observe that each $a_i$ only belongs to $\Sigma_{i}$ and $\Sigma_{i-1}$. Thus, $\mathcal{N}(a_i)=[1, n]-\{i-1, i\}$. We have the following case analysis:
\begin{enumerate}
\item $2<i_2-i_1<n-2$: In this case, the boundary symbols in $\Sigma_{i_1} \cup \Sigma_{i_2}$ are $a_{i_1}, a_{i_1+1}, a_{i_2}$ and $a_{i_2+1}$ (see Fig.~\ref{fig:ring}). 

a) For any symbol $a_j$ that lies between $a_{i_1+1}$ and $a_{i_2}$, 
we have $Cr_{\Delta}(\Sigma_{i_1} \cup \Sigma_{i_2}, a_j)=$ 
\begin{center}
$\overline{Cr}_{\Delta}(a_{i_1+1}a_{i_1+2}\ldots a_{j-2}a_{j-1}a_j) \cup \overline{Cr}_{\Delta}(a_{i_2}a_{i_2-1}\ldots a_{j+2}a_{j+1}a_j)$
\end{center}
It is not difficult to check that 
\begin{center}
$\overline{Cr}_{\Delta}(a_{i_1+1}a_{i_1+2}\ldots a_{j-2}a_{j-1}a_j)=[1, n]-\{i_1, i_1+1, i_1+2, \ldots, j-3, j-2, j-1\}$
\end{center}
and 
\begin{center}
$\overline{Cr}_{\Delta}(a_{i_2}a_{i_2-1}\ldots a_{j+2}a_{j+1}a_j)=[1, n]-\{j, j+1, j+2, \ldots, i_2-2, i_2-1, i_2\}$
\end{center}
Thus, $Cr_{\Delta}(\Sigma_{i_1} \cup \Sigma_{i_2}, a_j)=[1, n]$ for each $a_j$ between $a_{i_1+1}$ and $a_{i_2}$. 

b) For any symbol $a_j$ that lies between $a_{i_2+1}$ and $a_{i_1}$, 
we have $Cr_{\Delta}(\Sigma_{i_1} \cup \Sigma_{i_2}, a_j)=$ 
\begin{center}
$\overline{Cr}_{\Delta}(a_{i_2+1}a_{i_2+2}\ldots a_{j-2}a_{j-1}a_j) \cup \overline{Cr}_{\Delta}(a_{i_1}a_{i_1-1}\ldots a_{j+2}a_{j+1}a_j)$
\end{center}
It is not difficult to check that 
\begin{center}
$\overline{Cr}_{\Delta}(a_{i_2+1}a_{i_2+2}\ldots a_{j-2}a_{j-1}a_j)=[1, n]-\{i_2, i_2+1, i_2+2, \ldots, j-3, j-2, j-1\}$
\end{center}
and 
\begin{center}
$\overline{Cr}_{\Delta}(a_{i_1}a_{i_1-1}\ldots a_{j+2}a_{j+1}a_j)=[1, n]-\{j, j+1, j+2, \ldots, i_1-2, i_1-1, i_1\}$
\end{center}
Thus, $Cr_{\Delta}(\Sigma_{i_1} \cup \Sigma_{i_2}, a_j)=[1, n]$ for each $a_j$ between $a_{i_2+1}$ and $a_{i_1}$. Thus, if $2<i_2-i_1<n-2$, then for each $\sigma' \in \Sigma-(\Sigma_{i_1} \cup \Sigma_{i_2})$, $Cr_{\Delta}(\Sigma_{i_1} \cup \Sigma_{i_2}, \sigma')=[1, n]$
\item $1 \leq i_2-i_1\leq 2$: In this case, the boundary symbols in $\Sigma_{i_1} \cup \Sigma_{i_2}$ are $a_{i_1}$ and $a_{i_2+1}$. We only need to show that $Cr_{\Delta}(\Sigma_{i_1} \cup \Sigma_{i_2}, a_j)=[1, n]$ for each $a_j$ between $a_{i_2+1}$ and $a_{i_1}$. The proof is exactly the same as in  1) b. Thus, if $1 \leq i_2-i_1 \leq 2$, then for each $\sigma' \in \Sigma-(\Sigma_{i_1} \cup \Sigma_{i_2})$, $Cr_{\Delta}(\Sigma_{i_1} \cup \Sigma_{i_2}, \sigma')=[1, n]$ 
\item $n-2 \leq i_2-i_1\leq n-1$: In this case, the boundary symbols in $\Sigma_{i_1} \cup \Sigma_{i_2}$ are $a_{i_1+1}$ and $a_{i_2}$. We only need to show that $Cr_{\Delta}(\Sigma_{i_1} \cup \Sigma_{i_2}, a_j)=[1, n]$ for each $a_j$ between $a_{i_1+1}$ and $a_{i_2}$. The proof is exactly the same as in 1) a. Thus, if $n-2 \leq i_2-i_1\leq n-1$, then for each $\sigma' \in \Sigma-(\Sigma_{i_1} \cup \Sigma_{i_2})$, $Cr_{\Delta}(\Sigma_{i_1} \cup \Sigma_{i_2}, \sigma')=[1, n]$.

\hfill $\blacksquare$
\end{enumerate}

\begin{remark}
It is difficult to manage the proofs for results like Theorem~\ref{them: non-exi} by using $L_{cand}$ directly, since an infinite number of decomposability verifications would need to be carried out. The advantage of Theorem~\ref{theo: proof}, Corollaries~\ref{coro: simpli} and~\ref{coro:suffno} is that only structures of $\Delta$ and $\Delta_i$'s are required for the reasoning, using graphs of the dependence relations. This can facilitate reduction refutation for the parameterized setup, as illustrated in Theorem~\ref{them: non-exi}.
\end{remark}

It is not difficult to see that Proposition~\ref{theorem: char} and the candidate counter examples based reduction refutation technique are not comparable (see Example~\ref{examp: r5} and Example~\ref{examp: co}), and thus they are complementary for reduction refutation. 
\begin{example}
\label{examp: co}
Consider the distribution $\Delta=(\Sigma_1, \Sigma_2, \Sigma_3, \Sigma_4)$, where $\Sigma_1=\{a, b\}$, $\Sigma_2=\{b, c\}$, $\Sigma_3=\{c, d\}$ and $\Sigma_4=\{d, e\}$. Let $\{\Delta_1, \Delta_2\}$ be a candidate reduction of $\Delta$, where
\[   \left\{
\begin{array}{ll}
      \Delta_1=(\{a, b, c\}, \{c, d, e\}) \\
      \Delta_2=(\{a, b, c, d\}, \{d, e\}) 
\end{array} 
\right. \]
Proposition~\ref{theorem: char} can be used to easily refute $\{\Delta_1, \Delta_2\}$. However, $L_{cand}$ fails to refute $\{\Delta_1, \Delta_2\}$, since it is not decomposable with respect to $\Delta_1$, $\Delta_2$.
\end{example}
\vspace{-6pt}
\subsection{Strengthened Substitution-based Proof for Reduction Validation}
\label{sec: cr}
Let $P$ be any candidate reduction of $\Delta$. Suppose it passes the tests of Proposition~\ref{theorem: char} and the candidate counter example $L_{cand}$. Then, it is of interest to try validating $P$.
 A substitution-based proof technique has been proposed in~\cite{LWSW17b} for automated reduction validation. We here recall the main idea and provide a formal soundness proof. Then, we proceed to show that the substitution-based proof technique can be improved with the help of Lemma~\ref{prop:down}.

The main idea: for any distribution $\Delta'=(\Sigma'_1, \Sigma'_2, \ldots, \Sigma'_n)$ of $\Sigma$,  let $A(\Delta'):=\bigcup_{i \neq j}(\Sigma'_i \cap \Sigma'_j)$ denote the set of shared symbols in distribution $\Delta'$. Let $\Delta''=(\Sigma''_1, \Sigma''_2, \ldots, \Sigma''_l)$ be another distribution of $\Sigma$. $\Delta'$ is said to be {\em substitutable} into the $i$-th sub-alphabet of $\Delta''$, denoted $\Delta' \rightharpoondown_{i} \Delta''$, if $A(\Delta')\subseteq \Sigma''_i$ holds. $\Delta'$ is said to be {\em substitutable} into $\Delta''$, denoted $\Delta' \rightharpoondown \Delta''$, if $\Delta' \rightharpoondown_{i} \Delta''$ for some $i\in [1, l]$. The result of substituting $\Delta'$ into the $i$-th sub-alphabet of $\Delta''$ is denoted by $(\Delta' \vdash_i \Delta'')$, which is defined to be the distribution obtained from 
\vspace{-6pt}
\begin{center}
$(\Sigma''_1, \Sigma''_2, \ldots, \Sigma''_{i-1}, \Sigma''_i \cap \Sigma'_1, \Sigma''_i \cap \Sigma'_2, \ldots, \Sigma''_i \cap \Sigma'_n, \Sigma''_{i+1}, \ldots, \Sigma''_l)$
\end{center}
\vspace{-2pt}
by keeping only the maximal components. We denote the result of substituting $\Delta'$ into $\Delta''$  by $(\Delta' \vdash \Delta'')$, which is defined to be the set $\{(\Delta' \vdash_i \Delta'') \mid \Delta' \rightharpoondown_{i} \Delta'', i \in [1, l]\}$ of distributions. Given any candidate reduction $\{\Delta_1, \Delta_2, \ldots, \Delta_m\}$ of a source distribution $\Delta$, the idea of the proof technique~\cite{LWSW17b} is to derive new distributions by using substitutions and maintain the set of currently available distributions. If $\Delta$ can be obtained from $\{\Delta_1, \Delta_2, \ldots, \Delta_m\}$ in this way and also $\Delta \leq_{\Sigma} \Delta_i$ holds for each $i \in [1, m]$, then $\{\Delta_1, \Delta_2, \ldots, \Delta_m\}$ is a reduction of $\Delta$, using Lemma~\ref{lemm: separation} and Lemma~\ref{lemma:good}. In fact, let $\Delta'$ and $\Delta''$ be any two distributions of $\Sigma$, where $\Delta' \rightharpoondown_i \Delta''$. For any language $L$ over $\Sigma$, if $L$ is decomposable with respect to $\Delta'$ and $\Delta''$, then $L$ is also decomposable with respect to the distribution $(\Delta' \vdash_i \Delta'')$, using Lemma~\ref{lemm: separation}. 
\begin{lemma}
\label{lemma:substproo}
Let $\Delta', \Delta''$ be any two distributions of $\Sigma$, where $\Delta' \rightharpoondown_i \Delta''$. For any language $L$ over $\Sigma$, if $L$ is decomposable with respect to $\Delta'$ and $\Delta''$, then $L$ is also decomposable with respect to the distribution $(\Delta' \vdash_i \Delta'')$.
\end{lemma}
{\em Proof}:
By definition, $(\Delta' \vdash_i \Delta'')$ is the distribution obtained from 
\begin{center}
$(\Sigma''_1, \Sigma''_2, \ldots, \Sigma''_{i-1}, \Sigma''_i \cap \Sigma'_1, \Sigma''_i \cap \Sigma'_2, \ldots, \Sigma''_i \cap \Sigma'_n, \Sigma''_{i+1}, \ldots, \Sigma''_l)$
\end{center}
by keeping only the maximal components. If $L$ is decomposable with respect to $\Delta'$ and $\Delta''$, then we have
\[   \left\{
\begin{array}{ll}
      L=P_{\Sigma'_1}(L) \lVert P_{\Sigma'_2}(L) \lVert \ldots \lVert P_{\Sigma'_n}(L)  \hskip 6pt \ldots $(1)$\\
      L=P_{\Sigma''_1}(L) \lVert P_{\Sigma''_2}(L) \lVert \ldots \lVert P_{\Sigma''_l}(L) \hskip 4pt \ldots $(2)$\\
\end{array} 
\right. \]
By substituting $(1)$ to $(2)$, we have, by Lemma~\ref{lemm: separation}, 
\begin{center}
$L=P_{\Sigma''_1}(L) \lVert P_{\Sigma''_2}(L) \lVert \ldots \lVert P_{\Sigma''_{i-1}}(L) \lVert$ \\
$P_{\Sigma''_i}(P_{\Sigma'_1}(L) \lVert P_{\Sigma'_2}(L) \lVert \ldots \lVert P_{\Sigma'_n}(L)) \lVert$\\ $ P_{\Sigma''_{i+1}}(L) \lVert \ldots \lVert P_{\Sigma''_l}(L)=P_{\Sigma''_1}(L) \lVert P_{\Sigma''_2}(L) \lVert \ldots \lVert P_{\Sigma''_{i-1}}(L) \lVert$ \\
$P_{\Sigma''_i \cap \Sigma'_1}(L) \lVert P_{\Sigma''_i \cap \Sigma'_2}(L) \lVert \ldots \lVert P_{\Sigma''_i \cap \Sigma'_n}(L) \lVert$\\ $ P_{\Sigma''_{i+1}}(L) \lVert \ldots \lVert P_{\Sigma''_l}(L)$ 
\end{center}
For any term $P_{\Sigma'}(L)$ at the right hand side of the last equality, it can be removed, if there exists another term $P_{\Sigma''}(L)$ such that $\Sigma' \subseteq \Sigma'',$ and the equality still holds. Thus, we only need to keep those terms $P_{\Sigma''}(L)$ where $\Sigma''$ is maximal. 

We then conclude that $L$ is decomposable with respect to the distribution $(\Delta' \vdash_i \Delta'')$.
 \hfill $\blacksquare$\\

Thus, if $L$ is decomposable with respect to each $\Delta_i$, then the above substitution procedure would ensure that $L$ is also decomposable with respect to the newly generated distribution $\Delta$. If $\Delta \leq_{\Sigma} \Delta_i$ for each $i \in [1, m]$, then the decomposability of $L$ with respect to $\Delta$ also implies the decomposability of $L$ with respect to each $\Delta_i$, by Lemma~\ref{lemma:good}. We then conclude that $\{\Delta_1, \Delta_2, \ldots, \Delta_m\}$ is a reduction of $\Delta$. Thus, the substitution-based proof technique is sound. 

We now use the next example to show that the substitution-based proof technique is not complete, and it can be improved with the help of Lemma~\ref{prop:down}.
\begin{example}
\label{exa:sd}
Consider $\Delta=(\Sigma_1, \Sigma_2, \Sigma_3, \Sigma_4, \Sigma_5, \Sigma_6)$, where $\Sigma_1=\{a, b\}$, $\Sigma_2=\{b, c\}$, $\Sigma_3=\{a, c\}$, $\Sigma_4=\{d, e\}$, $\Sigma_5=\{e, f\}$ and $\Sigma_6=\{d, f\}$. Let $\{\Delta_1, \Delta_2\}$ be a candidate reduction of $\Delta$, where
\[   \left\{
\begin{array}{ll}
      \Delta_1=(\{a, b, d, e\}, \{b, c, e, f\}, \{a, c, d, f\}) \\
      \Delta_2=(\{a, b, c\}, \{d, e\}, \{e, f\}, \{d, f\}) 
\end{array} 
\right. \]
It is clear that the substitution-based proof cannot be applied on $\{\Delta_1, \Delta_2\}$. Furthermore, neither the necessary condition in Lemma~\ref{lemma: c} nor the candidate counter example $L_{cand}$ can refute $\{\Delta_1, \Delta_2\}$. Let $P=\{\Delta_1, \Delta_2\}$ and let $Q=\{\Delta_1, \Delta_2'\}$, where $\Delta_2'=(\{a, b, c\}, \{d, e, f\})$. It is clear that $\Delta_2 \leq_{\Sigma} \Delta_2'$ and thus $P \leq_{\Delta} Q$. It is possible to show, by using the substitution-based proof technique, that $Q$ is indeed a reduction of $\Delta$. Then, by Lemma~\ref{prop:down}, $P$ is also a reduction of $\Delta$. Thus, the substitution-based proof technique is not complete.
\end{example}

Based on Example~\ref{exa:sd}, we have the next observation. If the substitution-based proof technique cannot validate $P$ and the reduction refutation techniques fail to refute $P$, then we can check whether there exists a set $Q \subseteq \mathcal{M}(\Delta)$ of distributions such that $P \leq_{\Delta} Q$ and the substitution-based proof technique works on $Q$ (see Proposition~\ref{propos: subt}). $Q$ does not need to be an element in $S_{\Delta}$, that is, we do not require the distributions in $Q$ to be $\leq_{\Sigma}$-incomparable. The definition of $P \leq_{\Delta} Q$ can be naturally extended in this case: $\forall \Delta_j' \in Q, \exists \Delta_i \in P, \Delta_i \leq_{\Sigma} \Delta_j'$. In general, there are many different choices of $Q$. It turns out that we only need to apply the substitution-based proof on the following set 
\begin{center}
$\mathcal{U}(P):=\{\Delta' \in \mathcal{M}(\Delta) \mid \exists \Delta'' \in P, \Delta'' \leq_{\Sigma} \Delta'\}=\bigcup_{\Delta'' \in P}\mathcal{U}(\Delta'')$
\end{center}
of distributions, where $\mathcal{U}(\Delta''):=\{\Delta' \in \mathcal{M}(\Delta) \mid \Delta'' \leq_{\Sigma} \Delta''\}$. It is clear that $P \leq_{\Delta} \mathcal{U}(P)$ and any $Q \subseteq \mathcal{M}(\Delta)$ with $P \leq_{\Delta} Q$ is a subset of $\mathcal{U}(P)$. In general, $\mathcal{U}(P) \subseteq \mathcal{M}(\Delta)$ is not an element in $S_{\Delta}$. We immediately have the following.
\begin{proposition}
\label{propos: subt}
If $\Delta$ can be obtained from some $Q \subseteq \mathcal{M}(\Delta)$ (by applying the substitution-based proof on $Q$), where $P \leq_{\Delta} Q$ and $P \in S_{\Delta}$, then $P$ is a reduction of $\Delta$; moreover, if $\Delta$ can be obtained from some $Q \subseteq \mathcal{M}(\Delta)$ with $P \leq_{\Delta} Q$, then $\Delta$ can be obtained from $\mathcal{U}(P)$.
\end{proposition}

{\em Proof}: If $\Delta$ can be obtained from some $Q \subseteq \mathcal{M}(\Delta)$, then $Q$ is a reduction of $\Delta$. It is clear that, for any $Q \subseteq \mathcal{M}(\Delta)$, $Q$ is a reduction of $\Delta$ if and only if $[Q]$ is a reduction of $\Delta$. Thus, $[Q]$ is a reduction of $\Delta$. Now, $P \leq_{\Delta} Q$ implies that $P \leq_{\Delta} [Q]$. By Lemma~\ref{prop:down}, we conclude that $P$ is a reduction of $\Delta$. The second statement follows from the fact that any $Q \subseteq \mathcal{M}(\Delta)$ with $P \leq_{\Delta} Q$ is a subset of $\mathcal{U}(P)$. Thus, if $\Delta$ can be obtained from $Q$, then it can be obtained from the subset $Q$ of $\mathcal{U}(P)$. \hfill $\blacksquare$\\

We note that the above approach actually proves the stronger result that $Q$ or, equivalently, $[Q]$ is a reduction of $\Delta$. We call this new approach the strengthened substitution-based proof. However, it is more expensive to apply the substitution-based proof on $\mathcal{U}(P)$. In the worst case, the substitution-based proof has to be applied on the set $\mathcal{M}(\Delta)$ of all merged distributions, if $P=[\bot(\Delta)]$. Thus, in the implementation, we only apply the strengthened proof if the (ordinary) substitution-based proof fails to prove $P$. We remark that the substitution-based proof on $Q \subseteq \mathcal{M}(\Delta)$, where $P \leq_{\Delta} Q$, can be viewed as a heuristic for the proof on $\mathcal{U}(P)$ (by working on the subset $Q \subseteq \mathcal{U}(P)$). The same remark is also applicable to $P \subseteq \mathcal{U}(P)$.
\subsection{Reduction Verification Procedure}
\label{sub:RVP}
Up to now, we have provided some techniques for reduction validation and reduction refutation. In this subsection, we are ready to present the reduction verification procedure.

We shall now summarize the main steps of the overall procedure involved in reduction verification. Given any candidate reduction $P$ of $\Delta$, 
\begin{enumerate}
\item Check whether $\bigwedge_{\Delta_i \in P}\Delta_i=\Delta$ (Proposition~\ref{theorem: char}). If not, return NO.
\item Check whether the candidate counter example $L_{cand}$ is decomposable with respect to each distribution in $P$. If yes, then return NO.
\item Apply the substitution-based proof on $P$. If $\Delta$ is obtained, then return YES.
\item Apply the substitution-based proof on $\mathcal{U}(P)$, i.e., apply the strengthened substitution-based proof on $P$. If $\Delta$ is obtained, then return YES.
\item return UNKNOWN
\end{enumerate}
We can compute $\bigwedge_{\Delta_i \in P}\Delta_i$ incrementally, thanks to the property of associativity: $\bigwedge_{i=1}^l\Delta_i=(\ldots((\Delta_1 \wedge \Delta_2) \wedge \Delta_3) \ldots \wedge \Delta_{l-1})\wedge \Delta_l$. 
Step 2) can use the decomposability verification algorithm~\cite{WH91} or the structure-based technique in Theorem~\ref{theo: proof} (Theorem~\ref{theo: chac} in Appendix A is an improvement). We note that the first two steps constitute the reduction refutation procedure; the third and the forth steps constitute the reduction validation procedure.
 It is worth mentioning that the first four steps could be executed in parallel.

There are still many open problems that are left unanswered for the above reduction verification procedure. In particular, we do not know whether it is complete. Indeed, it is of interest to first address following open question.

{\bf Open Question}: The above reduction verification procedure works on $\mathcal{U}(P)$, in order to prove that $P$ is a reduction of $\Delta$. It is possible to come up with a proof scheme that is at least as powerful. Indeed, we can apply the (substitution-based) proof on 
\vspace{-3pt}
\begin{center}
$\mathcal{U}'(P)=\{\Delta' \in \Delta(\Sigma) \mid \exists \Delta'' \in P, \Delta'' \leq_{\Sigma} \Delta'\},$
\end{center}
\vspace{-3pt}
 where $\Delta(\Sigma)$ denotes the set of all the distributions of $\Sigma$. We notice that this also proves that $P$ is a reduction of $\Delta$, since $P \leq_{\Delta} \mathcal{U}'(P)$. Since $\mathcal{U}(P) \subseteq \mathcal{U}'(P)$, we can conclude that the proof scheme based on $\mathcal{U}'(P)$ is at least as powerful as the proof scheme based on $\mathcal{U}(P)$ (cf. Proposition~\ref{propos: subt}). We do not know whether it is strictly more powerful to work with $\mathcal{U}'(P)$. However, we note that the size of $\mathcal{U}'(P)$ is much larger than the size of $\mathcal{U}(P)$ in general, thus we have chosen to\footnote{Theoretically, we can work with $\mathcal{U}'(P)$ instead.} work with $\mathcal{U}(P)$ in Step 4).

\section{Towards Automatic Generation of Optimal Reductions}
\label{sec:agr}
In this section, we shall address the problem of computing small or (near-)optimal reductions of distributions. We present an incremental production algorithm in Section~\ref{section:pcr} for generating candidate reductions. When combined with the reduction validation procedure presented in Section~\ref{sub:RVP}, this allows for a rather fast generation of (near-)optimal reductions, if small reductions do exist. Then, a recursive algorithm for computing small reductions is shown in Section~\ref{sec: RPCR}.
\vspace{-8pt}
\subsection{Incremental Production of Candidate Reductions}
\label{section:pcr}
The problem of producing candidate reductions is a solved problem, if we do not worry about optimality of the reductions. Indeed, by Theorem~\ref{thm: exiten}, $[{\bot(\Delta)}]$ is already a  ``perfect" candidate reduction\footnote{Here, ``perfect" means that we do not need to look at the other candidate reductions: if $[{\bot(\Delta)}]$ is not a reduction of $\Delta$, then there is no reduction of $\Delta$.} of $\Delta$, if the search space $S_{\Delta}$ is non-empty. 
It is of interest, however, to consider the production of candidate reductions with smaller/the smallest width,  in order to compute (near-)optimal reductions. We shall now explain how Lemma~\ref{prop: nece} can be used for the efficient production of candidate reductions in an incremental manner (Incremental Collection), which may be followed by an incremental refinement procedure guided by Lemma~\ref{lemma: c} (Incremental Refinement).

\begin{figure}[t]
\centering
\hspace*{-1mm}
\includegraphics[width=1.4in, height=0.8in]{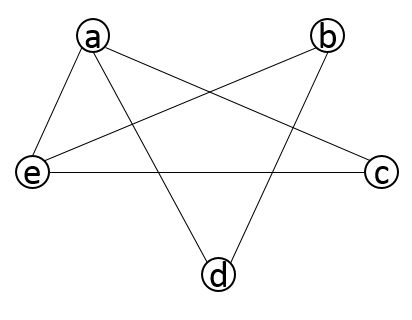} 
\caption{$I_{\Delta}$ induced by $\Delta=(\{a, b\}, \{b, c\}, \{c, d\}, \{d, e\})$}
\label{fig:inde3}
\end{figure}

By Lemma~\ref{prop: nece}, any reduction $\{\Delta_1, \Delta_2, \ldots, \Delta_l\}$ of $\Delta$ has to satisfy the condition $\bigcup_{i \in [1, l]}I_{\Delta_i}=I_{\Delta}$. To produce a candidate reduction, we incrementally collect pairs of independent symbols in $I_{\Delta}$, which correspond to edges in the graph $(\Sigma, I_{\Delta})$, using distributions $\Delta_i$'s that are merged from $\Delta$. We shall use the next example as an illustration of the idea.
\begin{example}
\label{exam: collect}
Consider the distribution $\Delta=(\Sigma_1, \Sigma_2, \Sigma_3, \Sigma_4)$ of $\{a, b, c, d, e\}$, where $\Sigma_1=\{a, b\}$, $\Sigma_2=\{b, c\}$, $\Sigma_3=\{c, d\}$, $\Sigma_4=\{d, e\}$. The independence relation $I_{\Delta}$ induced by $\Delta$ is represented in Fig.~\ref{fig:inde3}. By Lemma~\ref{prop: nece}, we need to produce a set $\{\Delta_1, \Delta_2, \ldots, \Delta_l\}$ of distributions such that $I_{\Delta}=\bigcup_{i \in [1, l]}I_{\Delta_i}$. Suppose we would like to collect the edge $\{a, e\}$ in $I_{\Delta}$ using $\Delta_1$, that is, ensure $\{a, e\} \in I_{\Delta_1}$ holds, then we only need to make sure that, in the process of merging $\Delta$, the sub-alphabets containing $a$ will not be merged with those sub-alphabets that contain $e$. It is not difficult to see that we have the following choices of $\Delta_1$ of size 2 (ignoring the ordering, which does not matter), which separates $\Sigma_1$ from $\Sigma_4$.
\begin{enumerate}
\item $\Delta_1=(\Sigma_1 \cup \Sigma_2 \cup \Sigma_3, \Sigma_4)$, or
\item $\Delta_1=(\Sigma_1 \cup \Sigma_2, \Sigma_4 \cup \Sigma_3)$, or
\item $\Delta_1=(\Sigma_1 \cup \Sigma_3, \Sigma_4 \cup \Sigma_2)$, or
\item $\Delta_1=(\Sigma_1, \Sigma_4 \cup \Sigma_2 \cup \Sigma_3)$
\end{enumerate} 
Suppose we choose option 1), then $\Delta_1=(\{a, b, c, d\}, \{d, e\})$. Then, $I_{\Delta_1}=\{\{a, e\}, \{b, e\}, \{c, e\}\}$. The remaining edges are then $I_{\Delta}-I_{\Delta_1}=\{\{a, c\}, \{a, d\}, \{b, d\}\}$. Suppose we would like to collect the edge $\{a, c\}$ in $I_{\Delta}-I_{\Delta_1}$ using $\Delta_2$, i.e., ensure $\{a, c\} \in I_{\Delta_2}$ holds, then we only need to make sure that, in the process of merging $\Delta$, the sub-alphabets containing $a$ are not merged with those sub-alphabets that contain $c$. It is not difficult to see that we have the following choices of $\Delta_2$ of size 2, which separates $\Sigma_1$ from $\Sigma_2$ and $\Sigma_3$.
\begin{enumerate}
\item [1')] $\Delta_2=(\Sigma_1 \cup \Sigma_4, \Sigma_2 \cup \Sigma_3)$, or
\item [2')] $\Delta_2=(\Sigma_1, \Sigma_2 \cup \Sigma_3 \cup \Sigma_4)$
\end{enumerate}
Suppose we choose option 2'), then $\Delta_2=(\{a, b\}, \{b, c, d, e\})$. Then, $I_{\Delta_2}=\{\{a, c\}, \{a, d\}, \{a, e\}\}$. The remaining edges are then $I_{\Delta}-I_{\Delta_1}-I_{\Delta_2}=\{\{b, d\}\}$. To collect edge $\{b, d\}$, we only have the following choice of $\Delta_3$ of size 2, which separates $\Sigma_1$ and $\Sigma_2$ from $\Sigma_3$ and $\Sigma_4$.
\begin{center}
$\Delta_3=(\Sigma_1 \cup \Sigma_2, \Sigma_3 \cup \Sigma_4)=(\{a, b, c\}, \{c, d, e\})$
\end{center}
Clearly, $\Delta_1, \Delta_2, \Delta_3$ are pairwise $\leq_{\Sigma}$-incomparable. Thus, we have produced a candidate reduction $\{\Delta_1, \Delta_2, \Delta_3\} \in S_{\Delta}$ that satisfies Lemma~\ref{prop: nece}. It is easy to check, using the substitution-based proof, that $\{\Delta_1, \Delta_2, \Delta_3\}$ is indeed a reduction of $\Delta$. 
\end{example}

To collect the edge $\{\sigma_1, \sigma_2\}$ in $I_{\Delta}$ using $\Delta_j$, we only need to make sure that those sub-alphabets that contain symbol $\sigma_1$ will not be merged with those sub-alphabets that contain $\sigma_2$. The Incremental Collection stage collects pairs of independent symbols using distributions $\Delta_i$'s in an incremental manner; it stops immediately after $\bigcup_{i \in [1, l]}I_{\Delta_i}=I_{\Delta}$ becomes satisfied. There are often different choices of distributions to be added at each step of the Incremental Collection stage. These decision choices form a tree structure; then, backtracking can be used to enumerate different choices. In order to facilitate the computation of small or (near-)optimal reductions, distributions with small sizes are explored first.

We note that the Incremental Collection stage allows for a quite efficient production of candidate reductions. However, it is still possible that the candidate reduction does not satisfy the necessary condition stated in Proposition~\ref{theorem: char}; the candidate reduction is then not a reduction of $\Delta$, and we must have $\bigwedge_{i=1}^l\Delta_i >\Delta$. If it holds that $\bigwedge_{\Delta' \in [{\bot(\Delta)}]}\Delta'=\Delta$, then it is guaranteed that we can collect more distributions in $\mathcal{M}(\Delta)$ so that the enlarged candidate reduction does meet the necessary condition. The enlarged candidate reduction can be considered a refinement of the original candidate reduction, which has a higher chance of becoming a reduction of $\Delta$ (c.f. Lemma~\ref{prop:down}). And, not surprisingly, Lemma~\ref{lemma: c} can be used for efficiently guiding the process of refining candidate reductions. The idea is first illustrated using the next example. For convenience, we also refer to the candidate reductions satisfying the necessary condition stated in Proposition~\ref{theorem: char} as {\em meet-consistent} candidate reductions.
\begin{example}
\label{example: inadd}
Consider $\Delta=(\Sigma_1, \Sigma_2, \Sigma_3, \Sigma_4, \Sigma_5, \Sigma_6)$, where $\Sigma_1=\{a, b\}$, $\Sigma_2=\{b, c\}$, $\Sigma_3=\{a, c\}$ and $\Sigma_4=\{d, e\}$, $\Sigma_5=\{e, f\}$, $\Sigma_6=\{d, f\}$. Suppose we merge $\Sigma_1$ and $\Sigma_2$, then we obtain the distribution $\Delta_1=(\{a, b, c\}, \{d, e\}, \{e, f\}, \{d, f\})$. It follows that $I_{\Delta_1}=I_{\Delta}$, that is, all the pairs of independent symbols of $\Delta$ have been collected by $\Delta_1$. Since the candidate reduction $\{\Delta_1\}$ is not meet-consistent, we could refine it by adding some more distributions that are merged from $\Delta$. Since $\{a, b, c\} \sqsubseteq \Delta_1$ and $\{a, b, c\} \not \sqsubseteq \Delta$, we can use $\Delta_2$ to break $\{a, b, c\}$, that is, ensure $\{a, b, c\} \not \sqsubseteq \Delta_2$. This would then help to ensure $\{a, b, c\} \not \sqsubseteq \Delta_1 \wedge \Delta_2$; and then we have $\Delta_1 \wedge \Delta_2=\Delta$, since $\{a, b, c\}$ is the only sub-alphabet in $\Delta_1$ that causes the trouble. To use $\Delta_2$ to break $\{a, b, c\}$, we must keep $\Sigma_1$, $\Sigma_2$ and $\Sigma_3$ separated in $\Delta_2$. Some possible choices of $\Delta_2$ (there are other choices that are not enumerated here) are 
\begin{enumerate}
\item $\Delta_2=(\{a, b\}, \{b, c\}, \{a, c\}, \{d, e, f\})$, or
\item $\Delta_2=(\{a, b, d, e, f\}, \{b, c\}, \{a, c\})$, or
\item $\Delta_2=(\{a, b\}, \{b, c, d, e, f\}, \{a, c\})$, or
\item $\Delta_2=(\{a, b\}, \{b, c\}, \{a, c, d, e, f\})$
\end{enumerate}
One can verify that the refined candidate reduction $\{\Delta_1, \Delta_2\}$ is indeed a reduction of $\Delta$, for any choice of $\Delta_2$ shown above, by using the substitution-based proof.
\end{example}
If the candidate reduction $\{\Delta_1, \Delta_2, \ldots, \Delta_l\}$ of distribution $\Delta=(\Sigma_1, \Sigma_2, \ldots, \Sigma_n)$ generated in the Incremental Collection stage does not satisfy the necessary condition $\bigwedge_{i=1}^l\Delta_i=\Delta$, then we shall need to break those troublesome sub-alphabets, according to Lemma~\ref{lemma: c}, by using distributions that are merged from $\Delta$ (in an incremental manner). Now, suppose $\Sigma' \subseteq \Sigma$ is some sub-alphabet that needs to be broken. Let $B(\Sigma')$ denote the minimal elements (in the sense of set inclusion) in  $\{I \subseteq [1, n] \mid \bigcup_{i \in I}\Sigma_i \supseteq \Sigma'\}$. We then add a distribution $\Delta'$ that is merged from $\Delta$, with the constraint that no element in $B(\Sigma')$ can be a subset of some element in the partition $P_{\Delta}(\Delta')$ corresponding to $\Delta'$. The same refinement procedure is repeated on the refined candidate reduction $\{\Delta_1, \Delta_2, \ldots, \Delta_l, \Delta'\}$. This Incremental Refinement process is continued until the refined candidate reduction becomes meet-consistent. There are often different choices of distributions to be added in each step of the refinement procedure. These decision choices form a tree structure, and backtracking can be used to enumerate different choices. Again, distributions with small sizes are explored first.

Of course, it is also possible to use the necessary condition in Lemma~\ref{lemma: c} to directly produce a candidate reduction, since collecting pairs of independent symbols is only a special case of breaking sub-alphabets, i.e., breaking sub-alphabets of size 2. And the procedure is the same as that of refining candidate reductions presented above, but starts with candidate reduction $\{\Delta_1\}$ instead. However, collecting pairs of independent symbols (with backtracking) alone is sufficient in most examples and works quite fast. This is illustrated using the next example.
\begin{example}
Let us now revisit Example~\ref{example: inadd}.
It is possible to generate a reduction in the Incremental Collection stage with a good choice of $\Delta_1$. With backtracking, we can first produce distribution $\Delta_1=(\{a, b\}, \{b, c\}, \{a, c, d, e, f\})$ instead. Since $I_{\Delta_1}$ is a proper subset of $I_{\Delta}$, we need more distributions. We can then add $\Delta_{2}=(\{a, b, c\}, \{d, e\}, \{e, f\}, \{d, f\})$ so that $I_{\Delta}=I_{\Delta_1} \cup I_{\Delta_2}$, since $I_{\Delta_2}=I_{\Delta}$. It is straightforward to verify that $\{\Delta_1, \Delta_2\}$ is a reduction of $\Delta$. Indeed, the same reduction has appeared in Example~\ref{example: inadd}, but in the reverse order. 
\end{example}
We now have the following remark regarding the Incremental Collection stage that collects pairs of independent symbols. 
\begin{remark}
Firstly, the order of production of distributions plays a useful role. To avoid the use of refinement of candidate reductions, one shall avoid $I_{\Delta_1}=I_{\Delta}$. Secondly, not all the reductions can be produced by collecting pairs of independent symbols. As a simple example, the reduction $\{\Delta_1, \Delta_2\}$ of distribution $\Delta$ in Example~\ref{example: inadd} cannot be generated by collecting pairs of independent symbols alone, where
\[   \left\{
\begin{array}{ll}
   \Delta_{1}=(\{a, b\}, \{b, c\}, \{a, c\}, \{d, e, f\}) \\
    \Delta_{2}=(\{a, b, c\}, \{d, e\}, \{e, f\}, \{d, f\}) \\
\end{array} 
\right. \]  
since $I_{\Delta_1}=I_{\Delta_2}=I_{\Delta}$. Thus, no matter whether $\Delta_1$ or $\Delta_2$ is produced first, the Incremental Collection stage will terminate there. It is clear that $\{\Delta_1, \Delta_2\}$ is not optimal, since $\{\Delta_1', \Delta_2'\}$ is an improved reduction of $\Delta$, where
\[   \left\{
\begin{array}{ll}
   \Delta_{1}'=(\{a, b, c\}, \{d, e, f\}) \\
    \Delta_{2}'=(\{a, b, d, e\}, \{b, c, e, f\}, \{a, c, d, f\}) \\
\end{array} 
\right. \]  
Fortunately, the improved reduction $\{\Delta_1', \Delta_2'\}$ can be generated by collecting pairs of independent symbols alone (with backtracking). It is of much interest to know whether there exist some distributions for which the Incremental Refinement stage is strictly necessary for the production of optimal reductions.
\end{remark}
A useful property for meet-consistent candidate reductions is the {\em minimality}. A meet-consistent candidate reduction $P$ of $\Delta$ is said to be {\em minimal} if every proper subset $P'$ of $P$ is not meet-consistent. 

We now summarize the main steps of the overall procedure involved in the automatic generation of small or (near-)optimal reductions using the incremental production approach. Given any distribution $\Delta$,
\begin{enumerate}
\item Check whether $\bigwedge_{\Delta' \in [{\bot(\Delta)}]}\Delta'=\Delta$. If not, return NO.
\item Check whether $L_{cand}$ is decomposable with respect to each distribution in $[{\bot(\Delta)}]$. If yes, return NO.
\item Generate a candidate reduction $P$ of $\Delta$ that has not been seen before, by collecting pairs of independent symbols (Incremental Collection stage). If $P$ does not exist and $[{\bot(\Delta)}]$ is minimal, then HALT.  If $P$ does not exist and $[{\bot(\Delta)}]$ is non-minimal, then goto Step 9). 
\item Check whether $\bigwedge_{\Delta_i \in P}\Delta_i=\Delta$. If yes, set $P':=P$, set $flag:=1$ and goto Step 6); otherwise, set $flag:=0$.
\item Refine candidate reduction $P$ by breaking sub-alphabets (Incremental Refinement stage) and generate a (refined) meet-consistent candidate reduction $P'$ that has not been seen before. If $P'$ does not exist, then goto Step 3).
\item Apply the substitution-based proof on $P'$ (Section~\ref{sec: cr}). If $\Delta$ is obtained, return $P'$
\item Apply the substitution-based proof on $\mathcal{U}(P')$, i.e., apply the strengthened substitution-based proof on $P'$. If $\Delta$ is obtained, return $P'$.
\item If $flag=1$, goto Step 3); otherwise, goto Step 5).
\item Apply the substitution-based proof on $[{\bot(\Delta)}]$. If $\Delta$ is obtained, return $[{\bot(\Delta)}]$
\item Apply the substitution-based proof on $\mathcal{M}(\Delta)$. If $\Delta$ is obtained, return $[{\bot(\Delta)}]$; otherwise, HALT.
\end{enumerate}
Step 1) and Step 2) are used to try refuting $[{\bot(\Delta)}]$; if $[{\bot(\Delta)}]$ is successfully refuted, then $\Delta$ does not have a reduction. The incremental production approach (with backtracking) consists of Step 3) to Step 10). We notice that this approach generates a minimal meet-consistent candidate reduction $P'$, before the (strengthened) substitution-based proof is applied on $P'$. If we cannot validate $P'$, then we backtrack and produce another minimal meet-consistent candidate reduction, if it exists. Since our purpose here is to generate reductions, we will give up on a candidate reduction $P'$ if the strengthened substitution-based proof fails on $P'$; in particular, we will not use $L_{cand}$ to refute $P'$. Backtracking (using tree structures) can be implemented for Step 3) and Step 5). In Step 4), $flag=1$ means that $P$ is meet-consistent. The procedure tests all the minimal meet-consistent candidate reductions in the worst case, if each $P'$ cannot be validated. If $[{\bot(\Delta)}]$ is minimal, then $[{\bot(\Delta)}]$ will be the last $P'$ that is produced. Then, in Step 3), the procedure terminates since we can neither refute $[{\bot(\Delta)}]$ nor validate it.

We still do not rule out the possibility that $\Delta$ has a reduction while every minimal meet-consistent candidate reduction (in the above sense) is not a reduction of $\Delta$. This is not a serious restriction, since we can apply the (strengthened) substitution-based proof on $[{\bot(\Delta)}]$ in case $[{\bot(\Delta)}]$ is non-minimal. Steps 9) and 10) are devoted just for that purpose. In particular, we have $\mathcal{U}([{\bot(\Delta)}])=\mathcal{M}(\Delta)$ in Step 10); Steps 9) and 10) will be executed if and only if $[{\bot(\Delta)}]$ is non-minimal and each minimal meet-consistent candidate reduction cannot be validated in Steps 3) to 8). We note that Steps 1), 2), 9) and 10) can be carried out in parallel and they can be run in parallel with Steps 3) to 8); then, the second part of Step 3) will be changed to ``If $P$ does not exist, then HALT", and $[\bot(\Delta)]$ does not need to be generated in Steps 3) to 8). We also need to ensure those distributions in a candidate reduction to be $\leq_{\Sigma}$-incomparable. This could be achieved by replacing $P$ and $P'$ with $[P]$ and $[P']$, respectively. We here shall note that every reduction $P'$ generated using the above approach, if it exists, is compact, due to the minimality.  

\begin{remark}
Unfortunately, we still cannot guarantee that the reduction generated by the above algorithm, if any, is optimal, since we cannot ensure that any candidate reduction that is not validated by the strengthened substitution-based proof, which is thrown away by the above algorithm, is indeed not valid. The algorithm only generates and tests upon minimal meet-consistent candidate reductions; we still do not rule out the possibility that a non-minimal meet-consistent candidate reduction can be optimal.  It is easy to modify the algorithm to collect the set of all validated reductions. Then, any optimal reduction among this collection can be chosen as the output.
\end{remark}

\vspace{-10pt}
\subsection{Recursive Generation of Reductions}
\label{sec: RPCR}
In this subsection, we shall develop a recursive algorithm that can be used for the production of small or (near-)optimal reductions. The idea is quite straightforward and is based on the next lemma.
\begin{lemma}
\label{lema: rr}
Let $P$ be any reduction of $\Delta$. Suppose some distribution $\Delta' \in P$ has a reduction $P'$.
 Then, $[P \cup P'-\{\Delta'\}]$ is also a reduction of $\Delta$.
\end{lemma}

 That is, suppose $P$ is a reduction of $\Delta$ and $P'$ is a reduction of some distribution $\Delta'$ in $P$. Then, we can replace $\Delta'$ with $P'$ and keep only those minimal distributions in $P \cup P'-\{\Delta'\}$. And the resulting set of distributions is still a reduction of $\Delta$. Then, Lemma~\ref{lema: rr} suggests a natural, ``conservative" algorithm for the recursive generation of (near-)optimal reductions.
The procedure thus shall proceed as follows: we first search for a reduction of $\Delta$ consisting of a small number of distributions in $[{\bot(\Delta)}]$. This can be easily carried out using the incremental production approach with backtracking, by ensuring that each distribution that is used to collect pairs of independent symbols is minimally merged from $\Delta$, followed by reduction validation. If such a reduction cannot be found, then we conclude that $\Delta$ does not have a reduction, by Theorem~\ref{thm: exiten}. Now, suppose that some reduction $P \subseteq [{\bot(\Delta)}]$ of $\Delta$ has been found. Then, we proceed to replace each distribution $\Delta' \in P$ using a reduction $P' \subseteq [{\bot(\Delta')}]$ of $\Delta'$, if it exists, as guided by Lemma~\ref{lema: rr}. The same procedure is carried out until no further replacement as above can be made. If there is a reduction of $\Delta$, then there usually exists a reduction of $\Delta$ that consists of a small number of distributions in $[{\bot(\Delta)}]$. It is usually much more efficient to verify these candidate reductions, since only a small number of substitutions are needed. 

There is an advantage of the recursive generation approach. If there exists a reduction, then it often can be confirmed after a few number of backtracking choices. The optimality of the final reduction depends on the initial reduction and the choices made in each replacement step, which can be enumerated with a tree structure. It is of interest to know whether all the optimal reductions can be computed in this way, which is still open. Again, there is no guarantee of the optimality for the generated reductions, since we are not sure about the completeness of the strengthened substitution-based proof.

\section{Conclusions}
\label{sec: condu}
In this work, we have further studied the notion of reduction of distributions and developed procedures for the computation of (near-)optimal reductions. In particular, we have procedures that can be used for the production of candidate reductions and reduction verification.
 There are still many technical problems that are left open. For example, we do not know whether the strengthened substitution-based proof for reduction validation is complete; and we still do not know whether the problem of reduction verification is decidable, although we conjecture that the answer is positive.

\vspace{-10pt}
\section*{Appendix A}

Theorem~\ref{theo: proof} is quite convenient to apply (cf. Corollary~\ref{coro:suffno}).
However, the condition stated in Theorem~\ref{theo: proof} is not necessary for the decomposability of $L_{cand}$ with respect to $\Delta'$. This shall be illustrated in the next example. Then, we will also explain how Theorem~\ref{theo: proof} can be improved.
\begin{example}
\label{exa: more}
Consider the distribution $\Delta=(\Sigma_1, \Sigma_2, \Sigma_3, \Sigma_4)$, where $\Sigma_1=\{a, b, c, g\}$, $\Sigma_2=\{c, d, e\}$, $\Sigma_3=\{d, e, f\}$ and $\Sigma_4=\{e, f, g\}$. $L_{cand}$ (the union of
\vspace{-2pt}
\begin{center}
$a \shuffle b \shuffle c \shuffle d^2 \shuffle e^2 \shuffle f^2 \shuffle g, a^3 \shuffle b^3 \shuffle c \shuffle d \shuffle e \shuffle f^3 \shuffle g^3, a^4 \shuffle b^4 \shuffle c^4 \shuffle d \shuffle e \shuffle f \shuffle g^4, a^5 \shuffle b^5 \shuffle c^5 \shuffle d^5 \shuffle e \shuffle f \shuffle g$)
\end{center}
\vspace{-2pt}
is indeed decomposable with respect to the distribution $\Delta'=(\{a, b, c, g\}, \{c, d, e, f\}, \{e, f, g\})$, which is obtained from $\Delta$ by merging $\Sigma_2$ and $\Sigma_3$. However, Theorem~\ref{theo: proof} fails to prove the decomposability of $L_{cand}$ with respect to $\Delta'$ in this case. Indeed, we notice that the sub-alphabet $\{c, d, e, f\}$ is the only element in $\Delta'-\Delta$; and the set $B_{\Delta'}(\{c, d, e, f\})$ of boundary symbols in $\{c, d, e, f\}$ is $\{c, e, f\}$. It is not difficult to check that $Cr_{\Delta'}(\{c, d, e, f\}, g)=[1, 4]$ and $Cr_{\Delta'}(\{c, d, e, f\}, a)=Cr_{\Delta'}(\{c, d, e, f\}, b)=\{2, 3, 4\} \neq [1, 4]$. With Theorem~\ref{theo: proof}, we only conclude that symbol $g$ is determined, and fail to prove the decomposability of $L_{cand}$ with respect to $\Delta'$. On the other hand, we observe that we have determined $\{c, d, e, f, g\}$ after symbol $g$ is determined. Thus, in particular, the numbers of occurrences of $c$ and $g$ are determined. We shall note that the only possibility is the following, which corresponds exactly to $L(\Sigma_j)$ for $j \in [1, 4]$
\begin{center}
$c \shuffle g$, $c \shuffle g^3$, $c^4 \shuffle g^4$ and $c^5 \shuffle g$.
\end{center}
Thus, the numbers of occurrences of $c$ and $g$ can determine the numbers of occurrences of $a$ and $b$ by matching $c$ and $g$ via $P_{\{a, b, c, g\}}(s_1)$. Thus, we have determined all the symbols in $\Sigma$ and $L_{cand}$ is decomposable with respect to $\Delta'$. We note that the reason $c$ and $g$ together can be used for matching is that they appear together only in the sub-alphabet $\Sigma_1$ of $\Delta$.
\end{example}
In the following, we propose an improved method for testing the decomposability of $L_{cand}$ with respect to distribution $\Delta' \in \mathcal{M}(\Delta)$. Given any proper sub-alphabet $\Sigma'$ of $\Sigma$, we compute $\mathcal{E}_{\Delta'}(\Sigma')$ with respect to $\Delta'$ as follows:
\begin{enumerate}
\item Set $\Sigma_d=\Sigma'$
\item Let $\Sigma_e$ denote $\{\sigma' \in \Sigma-\Sigma_d \mid Cr_{\Delta'}(\Sigma_d, \sigma')=[1, n]\}$. If $\Sigma_e=\varnothing$, then return $\Sigma_d$; otherwise, let $\Sigma_d:=\Sigma_d \cup \Sigma_e$.
\item If $\Sigma_d=\Sigma$, then return $\Sigma$; otherwise, let $\Sigma_{e'}:=$
\begin{center}
$\{\sigma' \in \Sigma-\Sigma_d \mid \exists \Sigma_d' \subseteq \Sigma_d, (|\Sigma_d'|\geq 2) \wedge (|\bigcup_{\sigma \in \Sigma_d'}\mathcal{N}(\sigma)|=n-1) \wedge (\{\sigma'\} \cup \Sigma_{d}' \sqsubseteq \Delta')\}$.
\end{center}
 If $\Sigma_{e'}=\varnothing$, then return $\Sigma_d$; otherwise, let $\Sigma_d:=\Sigma_d \cup \Sigma_{e'}$ and go to Step 2. 
\end{enumerate}
Intuitively, $\Sigma_d$ denotes the set of currently determined symbols at each step; $\Sigma_e$ denotes the set of symbols that are determined by matching chains of distinctive symbols (see Example~\ref{example:anally}); $\Sigma_{e'}$ denotes the set of symbols that are determined by matching two or more determined symbols in $\Sigma_d$ that appear together only in one sub-alphabet of $\Delta$ (see Example~\ref{exa: more}). We notice that, in Step 3), $|\bigcup_{\sigma \in \Sigma_d'}\mathcal{N}(\sigma)|=n-1$ holds if and only if the symbols in $\Sigma_d'$ appear together only in one sub-alphabet of $\Delta$; and then, with $\{\sigma'\} \cup \Sigma_{d}' \sqsubseteq \Delta'$, the symbols in $\Sigma_d'$ together can be used to determine $\sigma'$. It is clear that the above computation terminates; and we have the next results (for brevity, we omit the proofs due to space limitation).
\begin{theorem}
\label{theo: chac}
Let $\Delta'=(\Sigma_1', \Sigma_2', \ldots, \Sigma_k')$ be a distribution in $\mathcal{M}(\Delta)$. Then, $L_{cand}$ is decomposable with respect to $\Delta'$ if there exists a sub-alphabet $\Sigma_i' \in \Delta'-\Delta$ such that $\mathcal{E}_{\Delta'}(\Sigma_i')=\Sigma$, where $\Delta=(\Sigma_1, \Sigma_2, \ldots, \Sigma_n)$.
\end{theorem}

Theorem~\ref{theo: chac} is much more powerful than Theorem~\ref{theo: proof}. It is still an open problem whether the condition in Theorem~\ref{theo: chac} is necessary. 

 Let $\mathcal{E}_{\Delta}(\Sigma')$ be computed as in Step 1) to Step 3) of $\mathcal{E}_{\Delta'}(\Sigma')$, with the exception that $\Delta'$ is replaced with $\Delta$ throughout. We have the next corollary.
\begin{corollary}
\label{coro: gods}
Let $\Delta'=(\Sigma_1', \Sigma_2', \ldots, \Sigma_k')$ be a distribution in $\mathcal{M}(\Delta)$. Then, $L_{cand}$ is decomposable with respect to $\Delta'$ if there exists a sub-alphabet $\Sigma_i' \in \Delta'-\Delta$ such that $\mathcal{E}_{\Delta}(\Sigma_i')=\Sigma$, where $\Delta=(\Sigma_1, \Sigma_2, \ldots, \Sigma_n)$.
\end{corollary}

We have the next theorem, which can be used to establish stronger parameterized results.
\begin{theorem}
\label{theorem:neww}
$\Delta$ does not have a reduction if for any two different sub-alphabets $\Sigma_i, \Sigma_j \in \Delta$ such that $\Sigma_i \cup \Sigma_j \neq \Sigma$, it holds that $\mathcal{E}_{\Delta}(\Sigma_i \cup \Sigma_j)=\Sigma$, where $\Delta=(\Sigma_1, \Sigma_2, \ldots, \Sigma_n)$.
\end{theorem}

{\em Proof Sketch}: This immediately follows from Corollary~\ref{coro: gods} (cf. the proof of Corollary~\ref{coro:suffno}). \hfill $\blacksquare$ \\

\vspace{-16pt}
\section*{Appendix B}
We list some examples here for determining the existence of reductions for distributions, i.e., checking whether $[\bot(\Delta)]$ is a reduction of $\Delta$. In the third column, ``$L_{cand}$ (app)" means that $L_{cand}$ is a counter example for $[\bot(\Delta)]$ and the reasoning rule in Appendix A can be used; ``$L_{cand}$" means that $L_{cand}$ is a counter example and Theorem~\ref{theo: proof} can be used. ``sub" means that the (ordinary) substitution-based proof works for $[\bot(\Delta)]$. ``S sub" means that the strengthened substitution-based proof has to be used (and the ordinary proof fails). For distributions that have reductions, we show the wall-clock time for validating the existence of reductions using a non-optimized Python implementation of the strengthened proof on a laptop equipped with a 1.6GHz Intel Core i5-4200U CPU.  
\begin{center}
\begin{tabular}{ |c|c|c|c| } 
\hline
Distribution & Existence & Reason; Time \\
\hline
$(ab,bc,cd,dae)$ & no & $L_{cand}$ (app) \\ 
\hline
$(abe,bc,cd,daf)$ & no & $L_{cand}$ (app) \\ 
\hline
$(abe,bc,cde,da)$ & no & $L_{cand}$ \\ 
\hline
$(abe,bce,cde,dae)$ & no & $L_{cand}$ \\ 
\hline
$(abef,bce,cf,dae)$ & yes & sub; 0.08s \\ 
\hline
$(ab,bc,ac,df,de,ef)$ & yes & sub; 0.09s \\ 
\hline
$(abcf,abce,cdef,defg)$ & yes & sub; 0.10s \\ 
\hline
$(abcf,cde,def,cefg)$ & yes & sub; 0.08s \\ 
\hline
$(abg,bc,ac,df,de,efg)$ & yes &  S sub; 2.45s\\ 
\hline
$(abcg,cde,def,efg,fgca)$ & yes & S sub; 0.07s \\ 
\hline
$(abcg,cde,def,efg,cfgh)$ & yes & S sub; 0.29s \\ 
\hline
\end{tabular}
\end{center}

\end{document}